\documentclass[reprint,
 superscriptaddress,
 amsmath,amssymb,
 aps,
 prb,
]{revtex4-2}

\usepackage{bbold}
\usepackage{graphicx}
\usepackage{bm}
\usepackage{physics}

\usepackage{hyperref}
\hypersetup{
    colorlinks=true,
    linkcolor=blue,
    filecolor=blue,      
    urlcolor=blue,
    citecolor=blue
}

\usepackage{cleveref}

\begin{document}

\title{Constructing qubit edge states by inverse-designing the electromagnetic environment}

\author{A. Miguel-Torcal}
\email{alberto.miguel@uam.es}
\author{T. F. Allard}
\author{P. A. Huidobro}
\author{F. J. Garc\'ia-Vidal}
\author{A. I. Fern\'andez-Dom\'inguez}
\email{a.fernandez-dominguez@uam.es}
\affiliation{Departamento de F\'isica Te\'orica de la Materia Condensada, Universidad Aut\'onoma de Madrid, E- 28049 Madrid, Spain.}
\affiliation{Condensed Matter Physics Center (IFIMAC), Universidad Aut\'onoma de Madrid, E- 28049 Madrid, Spain.}



\date{\today}

\begin{abstract}
Building on advances in topological photonics and computational optimization, we inverse-design a periodic dielectric structure surrounding a chain of interacting qubits, emulating an extended, dimerized Su-Schrieffer-Heeger (SSH) excitonic model. Our approach enables precise control over photon-mediated interactions, allowing us to explore the emergence of topological edge states in the qubit chain. By systematically tuning structural parameters to address both coherent evolution and dissipative effects, we demonstrate that edge states remain robust and isolated from the bulk, even in the presence of long-range coupling and disorder, and preserving key topological properties despite deviations from complete chiral symmetry preservation. This work highlights the potential of inverse design in stabilizing topological excitonic states, opening new possibilities for advanced quantum technologies.
\end{abstract}

\maketitle


\section{Introduction}
The discovery of the integer quantum Hall effect in 1980~\cite{Klitzing1980} laid the groundwork for understanding topological phases in solid-state physics. This breakthrough led to the development of materials such as topological insulators~\cite{Kane2005,Zhang2006,Zhang2007}, which support surface states that are inherently protected from external noise and disorder. Inspired by these concepts, the field of topological photonics rapidly developed~\cite{Lu2014,Ozawa2019}, showing that similar states can also be engineered for photons~\cite{Haldane2008,Raghu2008}, resulting in unconventional behavior of light~\cite{Wang2008,Wang2009}. The flexibility and versatility of photonic systems, including experimental platforms like photonic crystals~\cite{Baile2022,Etxarri2020}, waveguides~\cite{Bello2019,Ke2016}, cavities and metasurfaces~\cite{Mann2022,Nie2020}, and QED circuits~\cite{Wang2016}, has facilitated the exploration of various exotic topological models and effects~\cite{StJean2017,Bandres2018}. This has offered new opportunities to create highly robust devices with applications in fault-tolerant quantum computation~\cite{Kitaev2003,Nayak2008,Field2018} and quantum state transfer~\cite{Angelis2020,Lang2017}.

A recent research avenue in advancing these practical implementations in quantum technologies focuses on exploring the interplay between topological systems and quantum emitters~\cite{Chang2018}, which can be realized with cold atoms~\cite{Bloch2017,Cooper2019,Broaweys2019} or molecules~\cite{Blackmore2019,Yicheng2023,Connor2023}, modeled as two-level systems (qubits). Similar works have considered localized bosonic modes, such as those sustained by subwavelength nanoparticles in the context of topological photonics~\cite{Kivshar2015,Wang2018,Downing2017}. Although it has been shown that qubit-based systems can exhibit symmetry-protected properties and robust excitonic edge states~\cite{Cirac2023},
the long-range nature of photon-mediated coupling in quantum optical platforms has proven to be a major challenge~\cite{Gong2016,Lepori2017}. Indeed, the impact of long-distance interactions on certain symmetries of the system influences the topological phase and edge states~\cite{LiChen2014,Platero2019,HsuChen2020,Faist2022}. 
The breaking of chiral symmetry in a 1D SSH model, driven by the strength of several hopping terms between qubits, can result in edge modes either retaining some level of protection while preserving certain topological properties or merging with the bulk bands, ultimately leading the system to metal-like behavior~\cite{Huidobro2019,Allard2023}. Moreover, as photonic environments undergo decoherence, not only can long-range coupling degrade topological properties, but non-Hermitian effects also tend to weaken topological protection, depending on the balance between dissipation and symmetry preservation~\cite{Nori2021,Gong2018,Svendsen2024}.

Controlling and tailoring long-range coupling is therefore crucial not only to benefit from adverse effects like decoherence but also to enhance topological robustness and enable new functionalities based on excitonic systems. In this regard, inverse design techniques~\cite{Molesky2018,BravoAbad2020} have been increasingly employed to optimize the interaction between qubits and their lossy environment, allowing for manipulating both coherent evolution and dissipation by systematically tuning structural parameters and material properties~\cite{Mignuzzi2019,Bennett2020,Bennett2021}. Particularly, topology optimization~\cite{Jensen2011} has been applied in this context enabling inverse-designed nanophotonic devices to achieve remarkable performance in qubit entanglement formation~\cite{AMT2022,AMT2024} and single-photon generation~\cite{Chakravarthi2020,Melo2023}. Despite having been successfully applied in topological photonics~\cite{Sigmund2019,Chen2020}, this approach has never been exploited to navigate the delicate trade-off between coherent and dissipative interaction mechanisms in excitonic systems to obtain stable and robust qubit topological states.

In this work, we investigate the emergence of topological edge states in quantum systems consisting of interacting qubits coupled to a periodic, inverse-designed dielectric structure. By means of a topology optimization strategy, we engineer a dielectric cavity (truncated waveguide) that accurately tunes the long-range photon-mediated exchange of excitations between qubits arranged in a wavelength-scale, equispaced chain, emulating the excitonic analog of the extended SSH model~\cite{SSH1979}. Crucially, unlike conventional topological-photonic implementations that rely on spatially varying inter-particle distances~\cite{Giannini2018} or orientations~\cite{Buendia2023}, our uniformly spaced qubit chain emulates an excitonic SSH model through the inverse design of the electromagnetic environment of the qubits. Our method operates in the single-excitation limit, in two stages: first addressing the topological properties of the Hamiltonian describing the coherent evolution of the system, and then focusing on the dissipative dynamics. We explore the effects of all-to-all hopping in our finite-size qubit chain and analyze its out-of-equilibrium dynamics. Our results reveal that edge states evolve in isolation from the bulk states, exhibiting decay and population exchange on a distinct timescale. Finally, we assess deviations from chiral symmetry preservation by examining the robustness of excitonic edge states against random disorder. Combining topological physics, quantum optics, and computational optimization, our results move the control and stability of topological qubit states forward, with potential applications in the next-generation of quantum devices.

\section{Physical setup and method}
The excitonic system we consider to achieve topological features consists of a collection of $N=12$ pairs of distant qubits, labeled A and B, with identical frequencies and perfect quantum yield. These are aligned along the same axis, resembling a one-dimensional lattice structure of $N$ bipartite unit cells. Importantly, both intra- and inter-cell spacings are uniform throughout the chain. The dynamics of the density matrix for the system, $\rho$, follows a master equation description of the form~\cite{Dung2002}
\begin{equation}
\frac{d\rho}{dt}=\frac{\imath}{\hbar}\Big[\rho,H\Big]+\sum_{i,j=1}^{2N}\gamma_{ij}\left(\sigma_j\rho\sigma_i^{\dagger}-\frac{1}{2}\left\lbrace\sigma_i^{\dagger}\sigma_j,\rho\right\rbrace\right),
\label{Eq1}
\end{equation}
under the assumption of weak coupling between the qubits and their electromagnetic environment. $\sigma_i$ ($\sigma^{\dagger}_i$) is the excitonic annihilation (creation) operator for the qubit located on site $i$ (ranging from 1 to 24), fulfilling anticommutation (commutation) relation $\lbrace\sigma_i,\sigma_i^{\dagger}\rbrace=1$ ($[\sigma_i,\sigma_j]=[\sigma_i,\sigma_j^{\dagger}]=0$ for $i\neq j$). The Hamiltonian in Eq.~\eqref{Eq1} can be written as 
\begin{equation}
H=\sum_{i=1}^{2N}\hbar\omega_0\sigma_i^\dagger\sigma_i+\sum_{\substack{i,j=1\\(i\neq j)}}^{2N}\hbar J_{ij}\sigma_i^\dagger\sigma_j,
\label{Eq2}
\end{equation}
where $\omega_0$ is the transition frequency of the qubits. The second term in Eq.~\eqref{Eq2} reflects the coherent interaction between the qubits mediated by off-resonant electromagnetic modes with strength given by $J_{ij}$. Referring back to Eq.~\eqref{Eq1}, the dissipative interaction between qubits ($i\neq j$) and their radiative decay ($i=j$), due to the coupling with on-resonant photonic modes, are incorporated through Lindblad operators weighted by the dissipative matrix (with entries $\gamma_{ij}$)~\cite{Petruccione2007}.   

Both coherent and dissipative coupling strengths are crucial to obtain the desired edge states and topological properties since both quantities determine the dynamics of edge modes through Equations \eqref{Eq1} and \eqref{Eq2}. These couplings are directly related to the dyadic Green's function of the qubits' electromagnetic environment~\cite{Novotny2012}. As shown by D\"ung \emph{et al.}~\cite{Dung2002}, this connection between the quantum dynamics of identical qubit ensembles and the spatial distribution of dielectric permittivity in their vicinity, $\epsilon(\vb r)$, is given by expressions $J_{ij}=\omega_0^{2}\vb{p}^*\real\{\vb{G}(\vb{r}_{i},\vb{r}_{j},\omega_0)\} \vb{p}/\hbar\varepsilon_{0}c^{2}$ and $\gamma_{ij}=2\omega_0^{2}\vb{p}^*\imaginary\{\vb{G}(\vb{r}_{i},\vb{r}_{j},\omega_0)\}\vb{p}/\hbar\varepsilon_{0}c^{2}$, where $\vb{p}$ is the transition dipole moment of the qubits, and $\vb{r}_{i,j}$ their position. Notably, the dyadic Green's function—and consequently the $J_{ij}$ and $\gamma_{ij}$ parameters—are evaluated at the qubits' natural frequency, $\hbar\omega_0\simeq 2.48$ eV ($\lambda=500$ nm). Our optimizing tool, focused on tailoring coherent and dissipative coupling strengths  through the dielectric environment, exploits precisely this dependence, which has enabled the investigation of qubit entanglement generation in various nanophotonic structures~\cite{GonzalezTudela2011,AMT2024}. Therefore, by tailoring the permittivity distribution of the hosting medium, we effectively control the master equation parameters to achieve resilient edge states.

We adapted the topology-optimization-based computational approach explained in detail in Ref.~\cite{AMT2022}, to first shape the coherent coupling strengths within the Hamiltonian, $J_{ij}$, tuning them to match the hopping amplitude distribution of the well-known dimerized SSH model. This adjustment should drive the system into a topological phase, enabling the emergence of edge modes-despite starting from an initially symmetric configuration of an equispaced qubit chain coupled through a homogeneous dielectric environment, namely free space. Subsequently, we manipulate the Lindblad terms through the dissipative matrix, $\gamma_{ij}$, to isolate their time evolution from that of bulk modes. This decoupling between the population dynamics of different Hamiltonian eigenstates may serve as evidence of the presence of edge states in our system, potentially signaling its topological nature. The numerical method employs a two-stage iterative procedure in which the real-valued permittivity map, $\epsilon(\vb{r})$, is progressively refined with a resolution set by the discretization employed in the solution of Maxwell’s equations. Starting from free space ($k=1$), each step ($k$) introduces a permittivity increment, $\var\epsilon$, for each mesh element and evaluates its impact on the corresponding target function, which first replicates the coherent coupling conditions characteristic of the topological phase in the SSH model and then reduces the radiative losses and dissipative interactions. An explicit expression for these target functions will be provided below. Only modifications that contribute towards its minimization are retained, shaping a dielectric cavity with the desired functionality. The algorithm’s speed and efficiency stem from first-order Born scattering series and Lorentz reciprocity, which streamline the evaluation of local dielectric variations on Dyadic Green’s functions~\cite{Novotny2012}. It is also worth noting that the design domain, and consequently the dielectric distribution, while optimized around the central pair of qubits in the chain (representing a single unit cell) is identically replicated across the remaining pairs. This constructs a finite periodic structure of bipartite unit cells that ultimately exhibits topological features.

\begin{figure}[tb!]
    \centering
    \includegraphics[width=\linewidth]{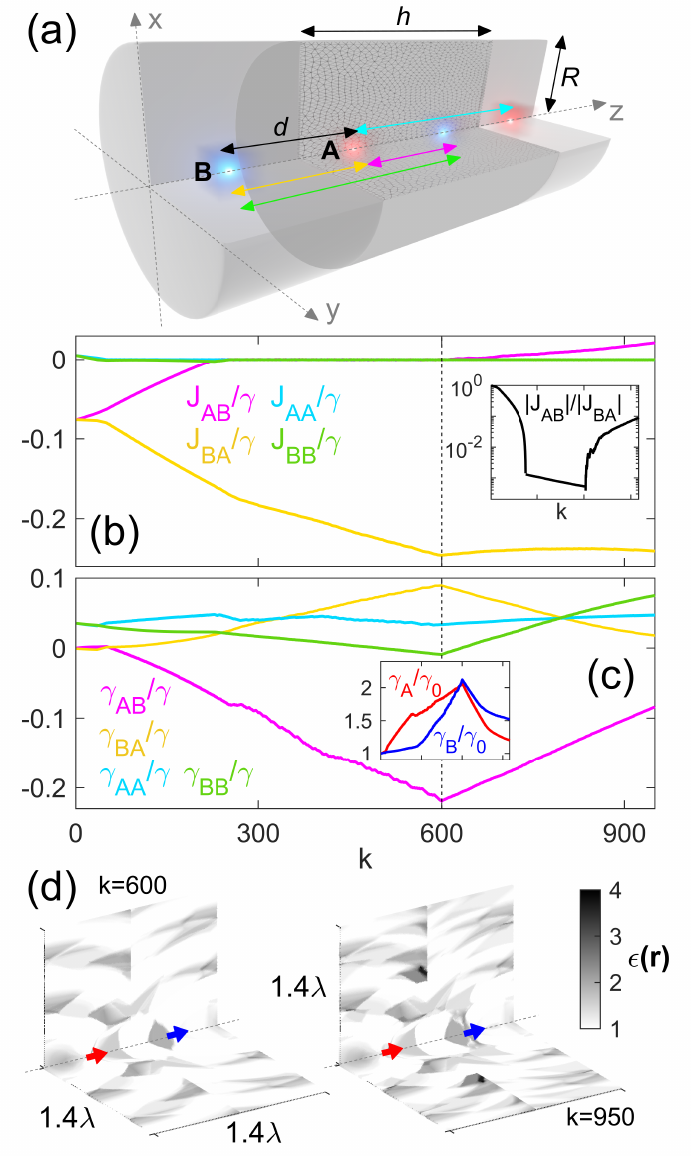}
    \caption{(a) Sketch of the system under study: A qubit chain with two sites per unit cell, named A and B, separated by a distance $d$, and a design domain with radius $R$ and unit cell size $h$ ($R=h=2d=716$ nm). (b) Qubit-qubit coherent interaction strengths as a function of the iteration step in the optimization procedure. The dashed line indicates the specific iteration step at which the target function is replaced. Inset: Coupling strength ratio between nearest neighbors inside and outside the unit cell. (c) First and second nearest neighbors dissipative coupling parameters versus iteration step. Inset: Decay rate normalized to vacuum radiation (Purcell factor) of each of the qubits inside the unit cell. (d) Unit cell of the inverse-designed dielectric structure at the end of each iterative stage. The relative permittivity continuously varies from 1 (white) to 4 (black).}
    \label{fig:1}
\end{figure}

Fig.~\ref{fig:1}(a) sketches our design space hosting the central qubit pair and its extension being reproduced on both sides of the chain. The optimization algorithm is integrated with the finite-element EM solver in Comsol Multiphysics$^{\rm TM}$, whose spatial discretization is represented by the dark gray thin mesh. The qubits are placed along the axial direction, with their dipole moments aligned parallel to it. Assuming a dipole moment of $\abs{\vb{p}}=1\,e\cdot$nm for the qubits~\cite{Eliseev2000}, this defines an energy scale set by the free-space decay rate $\hbar\gamma_0=\omega^3\abs{\vb{p}}^2/3\pi\epsilon_0 c^3=3.81\,\mu$eV. This longitudinal configuration both leverages the system’s azimuthal symmetry, allowing Maxwell’s equations to be solved in 2D, and facilitates the emergence of a topological phase in the SSH chain, as dipolar long-range couplings are less prominent than in the transverse case. Note that beyond-nearest-neighbour interactions between qubits belonging to the same sublattice (A or B) break chiral symmetry~\cite{Giannini2018}. As a result of our design procedure, we obtain a cylindrical waveguide with rotational symmetry, radius $R$, and unit cell size $h$, which is repeated throughout the qubit chain. The separation distance between the qubits, $d=358$ nm ($h=2d\approx1.4\lambda$), has been specifically chosen to cancel first-neighbor dissipative coupling in free space, $\imaginary\{\vb{G_0}(z_i,z_i+d,\omega_0)\}=0$~\cite{Vivas2021}, establishing an initially conducive configuration for the first stage of the optimization process (which focuses on the coherent sector of the system dynamics). For simplicity, the radius is also the same as unit cell size, $R=h=2d$.

\section{Results}
The iterative inverse design procedure of the dielectric cavity, which gives rise to edge states in the qubit chain’s single-excitation manifold, is analyzed in Figs.~\ref{fig:1}(b) and~\ref{fig:1}(c). Specifically, we examine the coupling strengths that govern photon-mediated interactions between qubits. As mentioned above, the optimization process proceeds in two distinct stages, separated by a dashed line at iteration $k=600$ in both panels. Each stage is guided by a different target function (see below), focusing on optimizing the coupling parameters related to either the coherent or dissipative terms of the master equation that describes the system's evolution. These parameters are scaled by the collective decay rate, $\gamma=\sqrt{\gamma_A\gamma_B}$~\cite{Ficek2002} (where $\gamma_A$ and $\gamma_B$ are the Purcell-enhanced decay rates for the central qubit pair in the presence of the inverse-designed dielectric structure), which serves as a reference energy scale throughout the entire study. Note that, due to the finite size of the system, relatively small spatial inhomogeneities exist in these decay rates, and in all the parameters in Eq.~\eqref{Eq1}, across the qubit chain. These variations are accounted for in our calculations, although, for simplicity, the discussion focuses on their values for the central qubits.  

In the first stage of the optimization process, the objective is to enhance inter-cell hopping, $\abs{J_{BA}}$, while reducing intra-cell hopping, $\abs{J_{AB}}$, thereby minimizing the ratio $\abs{J_{AB}}/\abs{J_{BA}}$. At the same time, the optimization suppresses second-nearest-neighbor couplings in the qubit chain, i.e. couplings between qubits referring to the same type of site (A or B) in neighboring unit cells, $\abs{J_{AA}}$ and $\abs{J_{BB}}$. As discussed above, these couplings play a crucial role, as they break chiral (sublattice) symmetry and hence the formal topological nature of the system. The exact expression for the target function in this first stage reads $f_1=\abs{J_{AB}}/\abs{J_{BA}}\times\abs{J_{AA}}\times\abs{J_{BB}}$, which is minimized during the topology optimization process. Note that the parameters concerning dissipative evolution remain unconstrained during the first part of the optimization procedure, as they are not included in the objective function. In contrast, from iteration $k=600$ onward, we replace the optimization function by one that releases the parameters referring to the coherent evolution and explicitly minimizes the elements of the dissipative matrix associated with radiative decay, $\gamma_A$ and $\gamma_B$, as well as the dissipative coupling between nearest neighbors, $\abs{\gamma_{AB}}$ and $\abs{\gamma_{BA}}$. Therefore, the target function for the second optimization stage takes the form $f_2=\gamma_A\times\gamma_B\times\abs{\gamma_{AB}}\times\abs{\gamma_{BA}}$. The coupling strengths are outlined in the sketch in Fig.~\ref{fig:1}(a) and depicted in Figs.~\ref{fig:1}(b) and~\ref{fig:1}(c) in the same color scale. For clarity, we note that the blue and green curves in panel (b) are superimposed.

The inset in Fig.~\ref{fig:1}(b) displays the ratio $\abs{J_{AB}}/\abs{J_{BA}}$ versus the iteration step, $k$, which must be minimized to resemble the nontrivial phase of an SSH chain with only nearest-neighbor interaction~\cite{Asboth2016}. Similarly, the inset in Fig.~\ref{fig:1}(c) illustrates the trend of the emission rates for qubits in the central unit cell. Since the structure repeats along the chain, the variations in emission rates between qubits are negligible, being slightly more prominent at the edges due to finite-size effects. The evolution of the parameters clearly reflects the abrupt change in the target function. During the initial stage, both $\abs{J_{AB}}/\abs{J_{BA}}$ ratio (black curve in the inset in panel (b)) and $\abs{J_{AA}}$ and $\abs{J_{BB}}$ rapidly shrink, regardless of the dissipation parameters, $\gamma_{ij}$. In the second stage, the parameters that control system losses are strongly modified, as evident from the sign change in their slopes, while those responsible for coherent evolution—already optimized—undergo little variation, except the ratio $\abs{J_{AB}}/\abs{J_{BA}}$. Although the latter increases, it stays smaller than unity and, as we will see in the following, allows the presence of edge states in the system. We also remark that coupling strengths vary significantly, while the collective radiative decay stays within 1 to 2 times $\gamma_0$, preserving the overall dynamical timescale (see inset in panel (c)). 

Fig.~\ref{fig:1}(d) shows the permittivity map, $\epsilon(\vb{r})$, of a unit cell of the dielectric waveguide at the final step of each optimization stage ($k=600$, left; $k=950$, right). A single unit cell is displayed, as it is periodically repeated along the chain. While the permittivity is fully defined within a single plane, it is presented in both the $xz$- and $yz$-planes for better visibility. The qubit pair is illustrated as blue and red arrows along the $z$-axis (representing the dipole momenta), and the gray scale represents the dielectric constant linearly varying from 1 (white) to its maximum (rather moderate) value at the end of the whole process, $\epsilon_{\rm max}=4$ (black). Both permittivity profiles exhibit elements characterized by intermediate dielectric constant values, clearly distinguishable from the surrounding free-space background. These elements extend from the region near the qubits to the waveguide edges. They influence the coherent interaction between the qubits by creating variations in each qubit’s environment, leading to a noticeable asymmetry between intra- and inter-cell couplings. Furthermore, as iterations increase and the system’s radiative losses are adjusted, more structure and new elements arise. Some of these elements eventually saturate, reaching the maximum permittivity value, which highlights a reduction in dissipative coupling strengths and a gradual decrease in each qubit’s radiative decay. Note that, despite this constraint, we did not apply any convergence criterion in either stage of our design procedure, and the transitions at the $600^{\rm th}$ and $950^{\rm th}$ iterations were set in a completely conventional manner.
\begin{figure}[tb!]
    \centering
    \includegraphics[width=\linewidth]{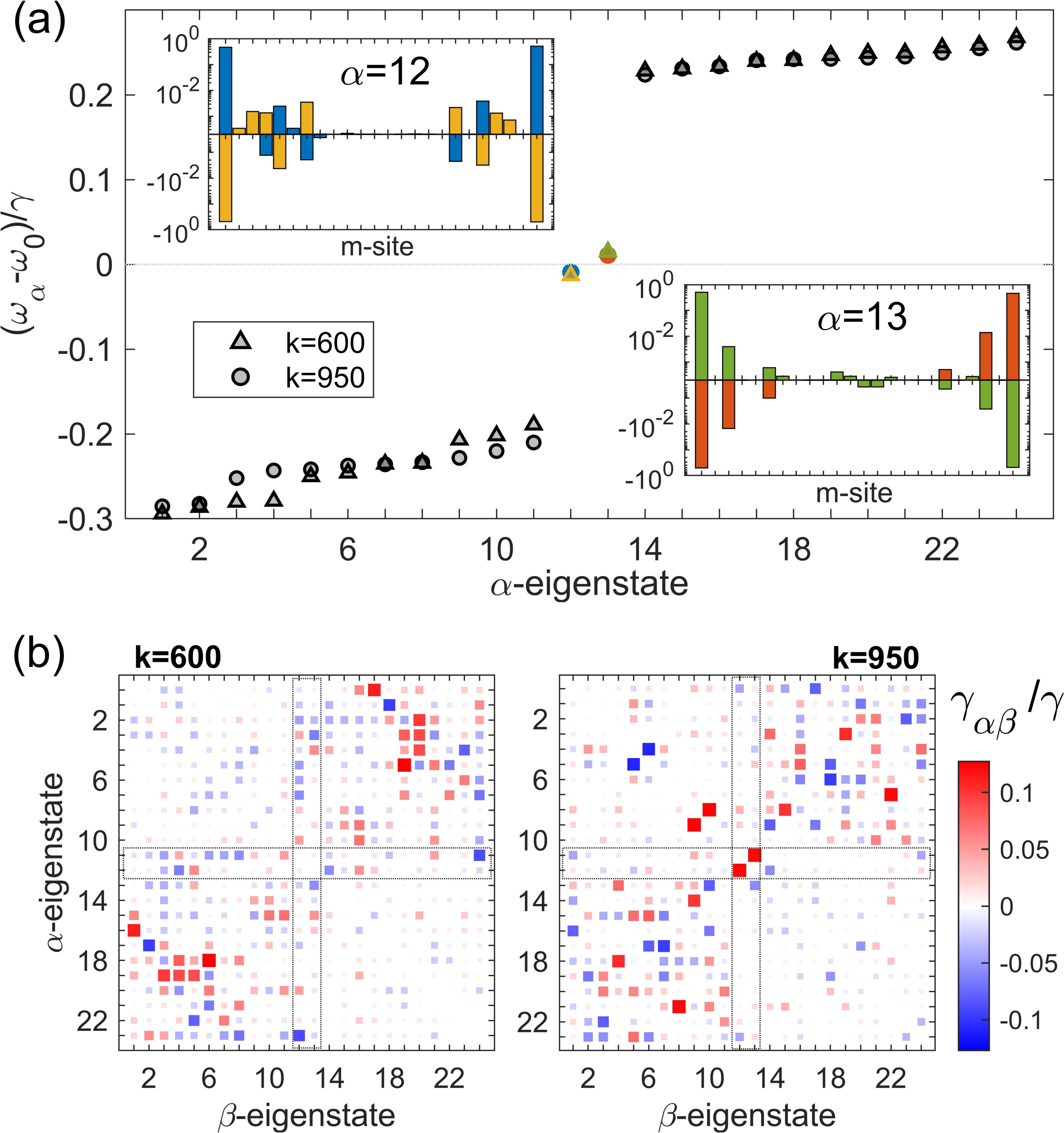}
    \caption{(a) Eigenstates of the Hamiltonian after each optimization stage ($k=600$, $k=950$) within the single excitation subspace. The real-valued wavefunction of the two edge-states that appear within the middle of the bandgap (indicated with blue and red circles, and yellow and green triangles) are represented in the insets in log scale, $\pm \rm sign(\phi_\alpha)\abs{\phi_\alpha}^2$ with $\alpha=12,13$. (b) Dissipative matrices in the Hamiltonian eigenstate basis at the end of the two optimization stages, normalized to the collective decay rate $\gamma=\sqrt{\gamma_A\gamma_B}$. Note that, for clarity, the diagonal terms (which correspond to the eigenstate decay rates) are not shown.}
    \label{fig:2}
\end{figure}

After analyzing the optimization process, we turn our attention to the edge modes in the excited-state population distribution of the qubit chain and the topological signatures they conceal. We therefore plot in Fig.~\ref{fig:2}(a) the real eigenvalues of the Hamiltonian in Eq. \eqref{Eq2}, in increasing order with $\alpha=1,2,...,24$, using the $J_{ij}$ parameters obtained at the last iteration step for the two target functions considered. Moreover, the insets display the eigenvectors corresponding to the two edge states, $\alpha=12,13$, by depicting their probability density as a function of the chain site, up to an arbitrary sign, $\pm \rm sign(\phi_\alpha)\abs{\phi_\alpha}^2$. This is introduced to retain phase information while increasing visibility of the two eigenstates (k=600, 950) in each panel. Note that these states are not localized at one edge of the chain, but appear at both ends, a feature discussed in more detail below. The eigenenergies in the main panel are rather symmetrically distributed around the qubits' transition frequency, $\omega_0$, forming two bands separated by a bandgap $\Delta\omega\equiv\omega_{14}-\omega_{11}\approx 0.4\gamma$. Within this bandgap, two edge states reside, well separated from the bulk bands—an effect reminiscent of topological systems in solid-state physics~\cite{Ando2013}. The bandgap size is comparable to, or somewhat smaller than, the characteristic energy scale associated with the collective emission rate, $\gamma=\sqrt{\gamma_A\gamma_B}$, as typically observed in photonic systems~\cite{Nori2021}. Notably, the long-range coherent coupling strengths between qubits on the same unit cell sites (A or B), induced by the dielectric environment, are small and weakly break chiral symmetry, leading to asymmetric energy bands and a slight energy separation between the edge states. After the second optimization stage, where dissipative interactions are suppressed, we observe a slightly increased bandgap, reduced band asymmetry, and a smaller energy separation between the edge states.

The eigenstates, shown in both insets of Fig.~\ref{fig:2}(a) with colors matching the symbols denoting their eigenenergies within the bandgap, exhibit strong edge-localized populations. These edge states, extended throughout the entire chain, form superpositions of A and B sites. In standard edge states with perfect overlap, the population concentrates in A (and B) sites at one end and in B (and A) sites at the opposite end. This population distribution, suggesting some chiral character, is most evident in the $\alpha = 13$ states. For $\alpha = 12$, an intensification of the chiral character can be inferred as the iteration progresses from $k=600$ to $k=950$. Moreover, the tuning of the dissipative coupling strengths during the second optimization stage indirectly affects the Hamiltonian, causing the eigenenergies of both states to approach degeneracy at the qubit transition frequency $\omega_0$.

We not only examine the presence of edge states and the emergence of topological signatures in the system through the parameters mediating coherent qubit-qubit coupling ($J_{ij}$), but also analyze how these interactions evolve dynamically and are influenced by system dissipation and decoherence. To this end, we compute the dissipative matrix in the eigenstate basis ($\gamma_{\alpha\beta}$) and track the edge state population decay and spatial redistribution. Fig.~\ref{fig:2}(b) shows this matrix, where color and size indicate dissipative coupling strengths between specific eigenstate pairs. The terms involving the midgap (edge) states are highlighted by a dashed rectangle. Also, to improve the visibility of the dissipative coupling between the two edge states and the rest of the first-excitation eigenstates (which from now on we term as bulk states), the decay rates (diagonal terms in $\gamma_{\alpha\beta}$) are not shown. The left panel ($k=600$) makes apparent that, before acting on the dissipative matrix, the dielectric cavity sustains well-defined states, localized at its edges, although they suffer a significant mixing with the bulk states due to decoherence effects in their dynamics. In the final design, after modifying the target function, the edge states become significantly more isolated from the bulk states and exhibit significantly stronger mutual coupling, as evidenced by the smaller dissipative couplings within the dashed rectangles in the right panel of Fig.~\ref{fig:2}(b), evaluated at $k=950$. This decoupling of the edge states from the bulk translates into their radiative decay into free space through an independent channel, which makes them a valuable resource for quantum technologies~\cite{Deng2021,Nie2021}.
\begin{figure}[tb!]
    \centering
    \includegraphics[width=\linewidth]{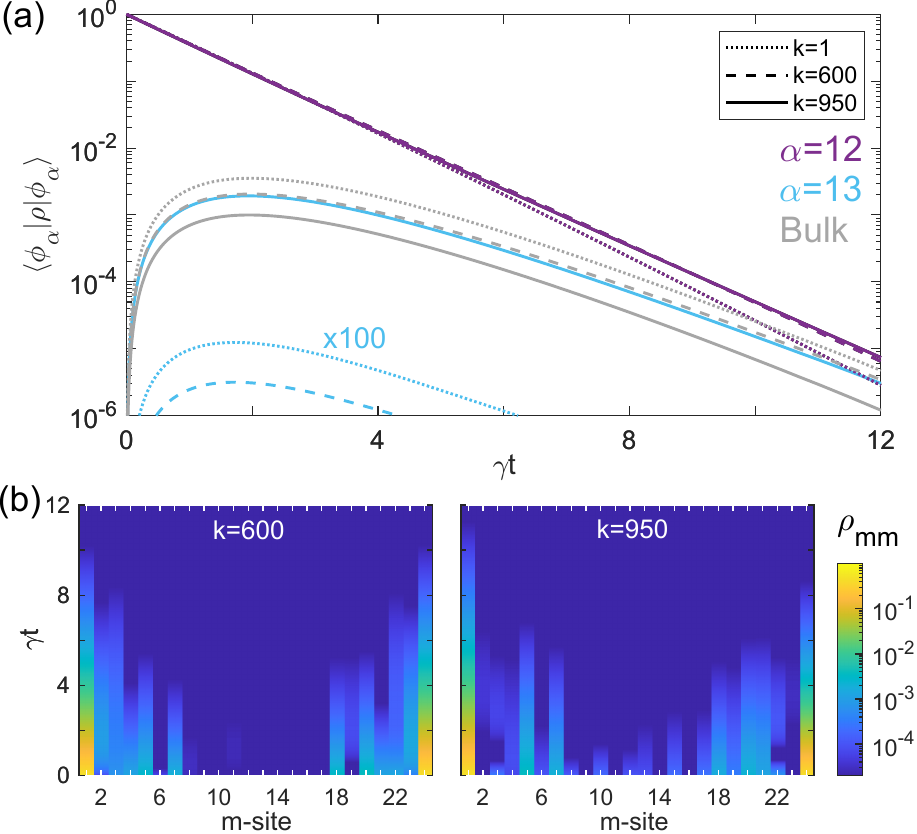}
    \caption{(a) Time evolution of the initially populated edge-state $\alpha=12$, both with and without the presence of the optimized dielectric structure, at the end of both optimization stages. (b) Population dynamics as a function of the chain site, $\rho_{mm}$, for $k=600$ (left) and $k=950$ (right), for the initially populated edge state, $\alpha=12$.} 
    \label{fig:3}
\end{figure}

To further analyze the edge state decoupled dynamics, we investigate next their evolution in time, by solving numerically Equation~\eqref{Eq1}. We initialize the system with one of the edge states fully populated, $\mel{\phi_{12}}{\rho(t=0)}{\phi_{12}}=1$, and let it decay spontaneously, rendering in Fig.~\ref{fig:3}(a) the population of this state and the other edge state, $\alpha=13$ (purple and light blue, respectively), along with the total population of the rest of the first-excitation eigenstates (gray), for both the free-space case ($k=1$) and the dielectric cavity after each optimization stage. We observe that, in the absence of the inverse-designed structure, the single excitation in the qubit chain decays within a moderately shorter time (see dotted purple line). During this process, the edge state does not only radiate into free space decaying into the ground state, but it also transfers population into the bulk states (dotted gray line). Note that their population is even larger than $\mel{\phi_{12}}{\rho(t)}{\phi_{12}}$ for $t\simeq 12/\gamma$. Additionally, in sufficiently long chains, the excitation remains completely decoupled from the other edge state, as the spatial separation between extreme sites and their different parity prevent significant interaction (dotted light blue line is scaled by a factor 100). As the dielectric cavity is engineered only at the Hamiltonian level ($k=600$), the decay of the initially populated edge state slows down due to an increased transfer to the other edge state, reducing its coupling with the bulk states (dashed lines). This effect becomes more pronounced after modifying the optimization function (solid lines, $k=950$), with the interaction between the two edge states becoming dominant over their interaction with the bulk eigenstates, despite the population being localized at the chain boundaries with a large separation between them (the edge sites are more than $16$ wavelengths apart in the chain).

Beyond the analysis of the eigenstate population dynamics, we examine next the edge state evolution in time and space by tracking the qubit population at the different sites of the chain. For this, we use the same initial state—the fully populated edge state $\alpha=12$—and plot the population at each site over time in the colored maps in Fig.~\ref{fig:3}(b), using a logarithmic scale. It is now clearer that the qubit population at the edges of the chain decays more slowly than the bulk sites, surviving above $t\simeq 6/\gamma$, when the population in the rest of the chain has vanished. As discussed above, the edge state evolution does not correspond to purely radiative decay into free space, but it is also affected by the interaction with other eigenstates of the system, which varies through the optimization procedure. In the site basis, and when optimizing for coherent evolution, only the edge qubits primarily exchange population with their nearest neighbors, with this interaction weakening over distance (left panel). In contrast, during the second stage, when the entries of the dissipative matrix are modified, intermediate qubits play a greater role, evolving even on a faster timescale, while edge-adjacent qubits become less populated (right panel). The bulk-decoupled spontaneous decay of edge states enables their detection and experimental verification through the light emitted into the far-field, benefiting from a qubit spacing in our inverse-designed truncated waveguide that is comparable to the natural wavelength.

The bulk topological properties of one-dimensional short-range quantum systems are captured by invariants such as the winding number~\cite{Asboth2016} and the Zak phase~\cite{Zak1989}. In the standard, nearest-neighbor SSH model, these bulk invariants, well-defined due to the system's inversion symmetry, enable the prediction of the presence of edge modes via the bulk-edge correspondence~\cite{Huidobro2019}. Moreover, these edge modes exhibit robustness against perturbations that preserve chiral symmetry, such as, e.g., disorder in the nearest-neighbor coupling constants. However, as mentioned in the introduction, the inclusion of long-range couplings—specifically, even-range interactions that connect sites of the same character (A or B)—breaks the chiral symmetry~\cite{Chiou2020}. This symmetry breaking changes the system’s topological class and invalidates the bulk-edge correspondence. Consequently, although edge states may still be present in our qubit chain, their formal topological protection is compromised due to the inherent complexities of the electromagnetic environment~\cite{McDonnell2022}. Nonetheless, we have already demonstrated that a carefully inverse-designed electromagnetic environment can still support strongly localized midgap edge states that dynamically evolve separated from the bulk. A key remaining question is whether these edge states obtained through our optimization process, although no longer topologically protected in a strict sense, retain some robustness to local disorder as a remnant of their topological origin.
\begin{figure}[bt!]
    \centering
    \includegraphics[width=\linewidth]{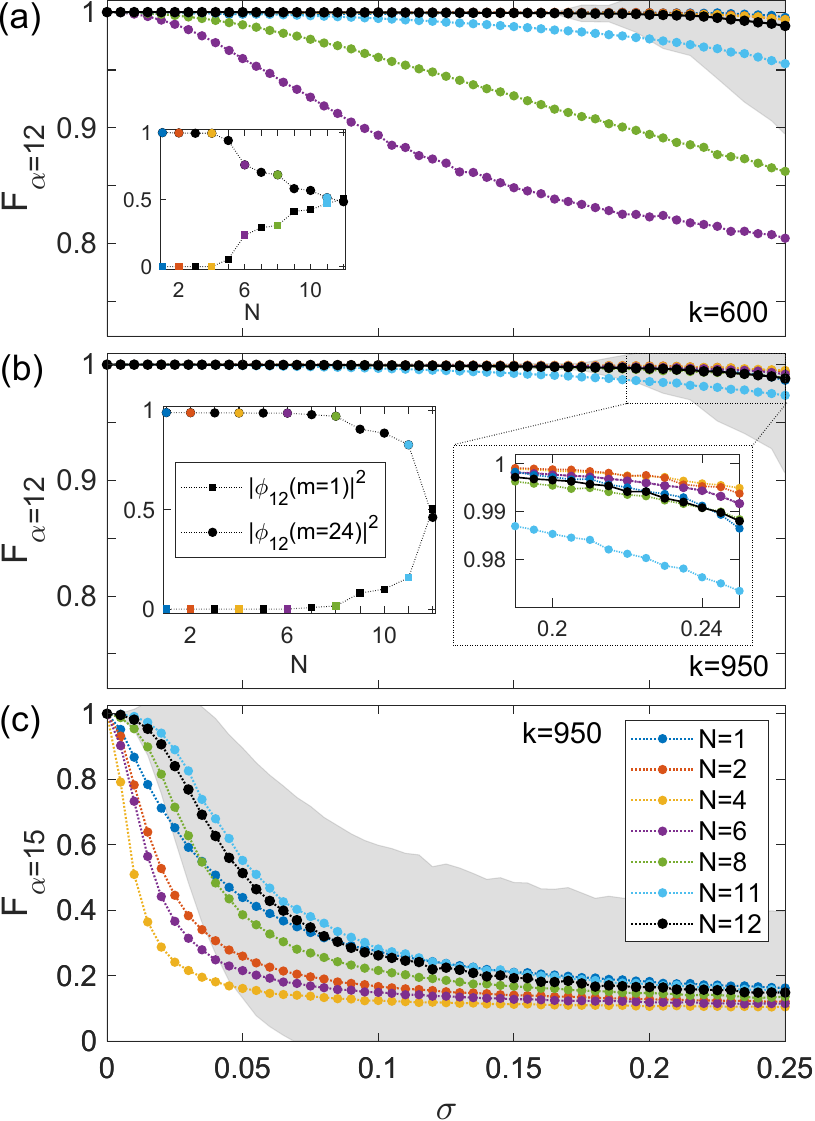}
    \caption{(a) Robustness of the edge states ($\phi_{12}$ at $k=600$) against normally distributed random noise introduced in the coherent coupling strengths and for different spatial ranges of the qubit-qubit interactions, indicated by $N$ (the number of neighboring unit cells in which coherent coupling strengths are non-zero). $F_\alpha$ is the fidelity between disordered and the original, non-disordered edge state. Inset: Population of the non-disordered edge-state $\phi_{12}$ at the chain's boundary sites. (b) Same as (a) but for the design obtained at $k=950$. (c) Same as (b) but for the bulk state $\alpha=15$.}
    \label{fig:4}
\end{figure}

To test the robustness of the edge states in our system (specifically $\alpha=12$), we introduce disorder in the coherent coupling parameters. Experimentally, such disorder can arise from unavoidable uncertainties in the positions of the qubits within the chain. We introduce random noise $\mathcal{R}$, according to a Gaussian distribution with standard deviation $\sigma$ and zero mean, centered on each coupling parameter in the Hamiltonian, $J_{ij}\rightarrow J_{ij}(1+\sigma \mathcal{R})$. The disorder level, controlled by $\sigma$, is scaled relative to the corresponding coupling strength, ensuring that the noise magnitude is proportional to the strength of the original interaction. For each realization of disorder, we compute the fidelity between the disordered and non-disordered edge states, $F_\alpha=\abs{\braket{\phi_\alpha^\sigma}{\phi_\alpha}}^2$~\cite{Jozsa1994}. We then average over multiple realizations, up to $2\cdot 10^4$, and present in Fig.~\ref{fig:4}(a) and 4(b) the mean fidelity (in black dots) as a function of the disorder strength for edge states after the first and second optimization stages, respectively. The standard deviation of $F_\alpha$ is shown in shaded gray areas. The panels also illustrate in different colors the mean fidelities between edge states in disordered and non-disordered systems as the interaction range extends from only nearest neighbors (equivalent to the standard SSH model, $N=1$) to all-to-all coupling ($N=12$). This is, we introduce a cut-off in the coherent coupling parameters, which are set to zero if the distance between sites is larger or equal to $N\cdot h$. This allows us to shed light into the effect of qubit-qubit interactions. In contrast, Fig.~\ref{fig:4}(c) shows the fidelity of a representative bulk state, such as $\alpha=15$ for $k=950$, as a function of disorder. The comparison against the upper panels makes evident the robustness of the edge states supported by our topology-optimized dielectric cavities surrounding the qubit chain.

We observe that the edge states remain highly protected against disorder and noise, losing only $1\%$ fidelity for $\sigma$ up to 0.25 across the entire finite chain. We interpret this observation as a fingerprint of the remnant topological character of the system, resulting from the weak breaking of chiral symmetry enabled by the suppression of long-range qubit interactions enabled by the inverse-designed dielectric cavity. In contrast, bulk states exhibit significantly greater fragility. We also note that the fidelity dependence on $N$ is non-monotonic, suggesting a strong dependence of the edge state wavefunction profile on the range of qubit interactions. Indeed, sensitivity to noise depends on the state localization at the edge sites. This is explored in the insets of Figs.~\ref{fig:4}(a) and~\ref{fig:4}(b), which plot the $\alpha=12$ population at the first ($m=1$) and last ($m=24$) sites of the chain. It reveals that for nearest-neighbor interactions ($N=1$), this state is completely localized at $m=1$. However, as the range of the photon-mediated coupling among qubits is increased, the state evolves and distributes equally at both edges of the chain. This variation of the edge state character, which we associate to finite size effects, explains the non-monotonic dependence on $F_{\alpha=12}$ on the noise level in the main panels. Remarkably, the robustness of the edge states under full-range interactions closely resembles that observed for $N=1$ in both designs.

By comparing the results of both optimization stages, we note that for the $k=950$ case, edge states are located closer to the mid-gap than in the $k=600$ case, as shown in Fig.~\ref{fig:2}(a), and their population becomes less mixed between the A and B sites. This suggests that the optimized structure at the end of the second stage is more chirally symmetric, and therefore, better protected against disorder. Indeed, minimizing dissipative coupling between neighboring qubits further stabilizes the edge states, enhancing robustness by counteracting the detrimental effects of long-range couplings. This leads to a reduction in the noise-induced standard deviation in the fidelities from Fig.~\ref{fig:4}(a) to~\ref{fig:4}(b), as indicated by the gray shaded area around the mean value. Finally, in the Supporting Information, we present the cavity design obtained by means of a single-stage optimization process using a unified target function that involves both coherent and dissipative coupling strengths. Due to the inherent complexity of the function, the master equation parameters saturate to values which yield a worse performance than the devices described here.

\section{Conclusions}
In this study, we have engineered robust edge states within the single-excitation manifold of a qubit chain by tailoring its electromagnetic environment. Through a topology optimization approach, we designed a dielectric cavity (truncated waveguide) that simultaneously controls photon-mediated interactions governing energy exchange between qubits and their radiative decay into free-space to enhance the system's topological properties. First, we have conducted a detailed analysis of the effectiveness of the inverse-designed devices in generating edge states by tracking the evolution of coupling parameters throughout the iterative optimization process. Next, we have investigated the excitonic edge states hosted in our system, examining their spatial distribution and symmetry properties. Their time evolution has then been explored by solving the population dynamics in the qubit chain with the cavity in place, proving that their spontaneous decay primarily involves edge-localized qubits and is decoupled from the dynamics of the rest of the (bulk) states. Finally, we have characterized the sensitivity of our edge states against potential disorder, demonstrating how their topological origin confers robustness. Notably, we have shown that both coherent and dissipative couplings contribute to protect them, even in the presence of significant noise. We are confident that our findings highlight inverse design as an effective approach for developing, optimizing and advancing topological features in quantum hardware based on nanophotonic platforms.

\begin{acknowledgments}
We acknowledge financial support from the Spanish Ministry of Science, Innovation and Universities through Grants PID2021-126964OB-I00, PID2021-125894NB-I00, PID2022-141036NAI00 (financed by MCIN/AEI/10.13039/501100011033 and FSE+), RYC2021-031568-I and CEX2018-000805-M, the European Union through ERC grant TIMELIGHT (grant agreement 101115792), and the European Union’s Horizon Europe Research and Innovation Programme through grant agreement No. 101070700.
\end{acknowledgments}


\section*{Data Availability Statement}
The data that support the findings of this study, including simulation templates and scripting codes developed in COMSOL Multiphysics and MATLAB, are available from the corresponding authors upon reasonable request. 
 
\bibliography{main}

\begin{thebibliography}{76}%
\makeatletter
\providecommand \@ifxundefined [1]{%
 \@ifx{#1\undefined}
}%
\providecommand \@ifnum [1]{%
 \ifnum #1\expandafter \@firstoftwo
 \else \expandafter \@secondoftwo
 \fi
}%
\providecommand \@ifx [1]{%
 \ifx #1\expandafter \@firstoftwo
 \else \expandafter \@secondoftwo
 \fi
}%
\providecommand \natexlab [1]{#1}%
\providecommand \enquote  [1]{``#1''}%
\providecommand \bibnamefont  [1]{#1}%
\providecommand \bibfnamefont [1]{#1}%
\providecommand \citenamefont [1]{#1}%
\providecommand \href@noop [0]{\@secondoftwo}%
\providecommand \href [0]{\begingroup \@sanitize@url \@href}%
\providecommand \@href[1]{\@@startlink{#1}\@@href}%
\providecommand \@@href[1]{\endgroup#1\@@endlink}%
\providecommand \@sanitize@url [0]{\catcode `\\12\catcode `\$12\catcode `\&12\catcode `\#12\catcode `\^12\catcode `\_12\catcode `\%12\relax}%
\providecommand \@@startlink[1]{}%
\providecommand \@@endlink[0]{}%
\providecommand \url  [0]{\begingroup\@sanitize@url \@url }%
\providecommand \@url [1]{\endgroup\@href {#1}{\urlprefix }}%
\providecommand \urlprefix  [0]{URL }%
\providecommand \Eprint [0]{\href }%
\providecommand \doibase [0]{https://doi.org/}%
\providecommand \selectlanguage [0]{\@gobble}%
\providecommand \bibinfo  [0]{\@secondoftwo}%
\providecommand \bibfield  [0]{\@secondoftwo}%
\providecommand \translation [1]{[#1]}%
\providecommand \BibitemOpen [0]{}%
\providecommand \bibitemStop [0]{}%
\providecommand \bibitemNoStop [0]{.\EOS\space}%
\providecommand \EOS [0]{\spacefactor3000\relax}%
\providecommand \BibitemShut  [1]{\csname bibitem#1\endcsname}%
\let\auto@bib@innerbib\@empty
\bibitem [{\citenamefont {Klitzing}\ \emph {et~al.}(1980)\citenamefont {Klitzing}, \citenamefont {Dorda},\ and\ \citenamefont {Pepper}}]{Klitzing1980}%
  \BibitemOpen
  \bibfield  {author} {\bibinfo {author} {\bibfnamefont {K.~v.}\ \bibnamefont {Klitzing}}, \bibinfo {author} {\bibfnamefont {G.}~\bibnamefont {Dorda}},\ and\ \bibinfo {author} {\bibfnamefont {M.}~\bibnamefont {Pepper}},\ }\bibfield  {title} {\bibinfo {title} {New method for high-accuracy determination of the fine-structure constant based on quantized hall resistance},\ }\href {https://doi.org/10.1103/PhysRevLett.45.494} {\bibfield  {journal} {\bibinfo  {journal} {Phys. Rev. Lett.}\ }\textbf {\bibinfo {volume} {45}},\ \bibinfo {pages} {494} (\bibinfo {year} {1980})}\BibitemShut {NoStop}%
\bibitem [{\citenamefont {Kane}\ and\ \citenamefont {Mele}(2005)}]{Kane2005}%
  \BibitemOpen
  \bibfield  {author} {\bibinfo {author} {\bibfnamefont {C.~L.}\ \bibnamefont {Kane}}\ and\ \bibinfo {author} {\bibfnamefont {E.~J.}\ \bibnamefont {Mele}},\ }\bibfield  {title} {\bibinfo {title} {${Z}_{2}$ topological order and the quantum spin hall effect},\ }\href {https://doi.org/10.1103/PhysRevLett.95.146802} {\bibfield  {journal} {\bibinfo  {journal} {Phys. Rev. Lett.}\ }\textbf {\bibinfo {volume} {95}},\ \bibinfo {pages} {146802} (\bibinfo {year} {2005})}\BibitemShut {NoStop}%
\bibitem [{\citenamefont {Bernevig}\ \emph {et~al.}(2006)\citenamefont {Bernevig}, \citenamefont {Hughes},\ and\ \citenamefont {Zhang}}]{Zhang2006}%
  \BibitemOpen
  \bibfield  {author} {\bibinfo {author} {\bibfnamefont {B.~A.}\ \bibnamefont {Bernevig}}, \bibinfo {author} {\bibfnamefont {T.~L.}\ \bibnamefont {Hughes}},\ and\ \bibinfo {author} {\bibfnamefont {S.-C.}\ \bibnamefont {Zhang}},\ }\bibfield  {title} {\bibinfo {title} {Quantum spin hall effect and topological phase transition in hgte quantum wells},\ }\href {https://doi.org/10.1126/science.1133734} {\bibfield  {journal} {\bibinfo  {journal} {Science}\ }\textbf {\bibinfo {volume} {314}},\ \bibinfo {pages} {1757} (\bibinfo {year} {2006})}\BibitemShut {NoStop}%
\bibitem [{\citenamefont {König}\ \emph {et~al.}(2007)\citenamefont {König}, \citenamefont {Wiedmann}, \citenamefont {Brüne}, \citenamefont {Roth}, \citenamefont {Buhmann}, \citenamefont {Molenkamp}, \citenamefont {Qi},\ and\ \citenamefont {Zhang}}]{Zhang2007}%
  \BibitemOpen
  \bibfield  {author} {\bibinfo {author} {\bibfnamefont {M.}~\bibnamefont {König}}, \bibinfo {author} {\bibfnamefont {S.}~\bibnamefont {Wiedmann}}, \bibinfo {author} {\bibfnamefont {C.}~\bibnamefont {Brüne}}, \bibinfo {author} {\bibfnamefont {A.}~\bibnamefont {Roth}}, \bibinfo {author} {\bibfnamefont {H.}~\bibnamefont {Buhmann}}, \bibinfo {author} {\bibfnamefont {L.~W.}\ \bibnamefont {Molenkamp}}, \bibinfo {author} {\bibfnamefont {X.-L.}\ \bibnamefont {Qi}},\ and\ \bibinfo {author} {\bibfnamefont {S.-C.}\ \bibnamefont {Zhang}},\ }\bibfield  {title} {\bibinfo {title} {Quantum spin hall insulator state in hgte quantum wells},\ }\href {https://doi.org/10.1126/science.1148047} {\bibfield  {journal} {\bibinfo  {journal} {Science}\ }\textbf {\bibinfo {volume} {318}},\ \bibinfo {pages} {766} (\bibinfo {year} {2007})}\BibitemShut {NoStop}%
\bibitem [{\citenamefont {Lu}\ \emph {et~al.}(2014)\citenamefont {Lu}, \citenamefont {Joannopoulos},\ and\ \citenamefont {Soljačić}}]{Lu2014}%
  \BibitemOpen
  \bibfield  {author} {\bibinfo {author} {\bibfnamefont {L.}~\bibnamefont {Lu}}, \bibinfo {author} {\bibfnamefont {J.~D.}\ \bibnamefont {Joannopoulos}},\ and\ \bibinfo {author} {\bibfnamefont {M.}~\bibnamefont {Soljačić}},\ }\bibfield  {title} {\bibinfo {title} {Topological photonics},\ }\href {https://doi.org/10.1038/nphoton.2014.248} {\bibfield  {journal} {\bibinfo  {journal} {Nature Photonics}\ }\textbf {\bibinfo {volume} {8}},\ \bibinfo {pages} {821} (\bibinfo {year} {2014})}\BibitemShut {NoStop}%
\bibitem [{\citenamefont {Ozawa}\ \emph {et~al.}(2019)\citenamefont {Ozawa}, \citenamefont {Price}, \citenamefont {Amo}, \citenamefont {Goldman}, \citenamefont {Hafezi}, \citenamefont {Lu}, \citenamefont {Rechtsman}, \citenamefont {Schuster}, \citenamefont {Simon}, \citenamefont {Zilberberg},\ and\ \citenamefont {Carusotto}}]{Ozawa2019}%
  \BibitemOpen
  \bibfield  {author} {\bibinfo {author} {\bibfnamefont {T.}~\bibnamefont {Ozawa}}, \bibinfo {author} {\bibfnamefont {H.~M.}\ \bibnamefont {Price}}, \bibinfo {author} {\bibfnamefont {A.}~\bibnamefont {Amo}}, \bibinfo {author} {\bibfnamefont {N.}~\bibnamefont {Goldman}}, \bibinfo {author} {\bibfnamefont {M.}~\bibnamefont {Hafezi}}, \bibinfo {author} {\bibfnamefont {L.}~\bibnamefont {Lu}}, \bibinfo {author} {\bibfnamefont {M.~C.}\ \bibnamefont {Rechtsman}}, \bibinfo {author} {\bibfnamefont {D.}~\bibnamefont {Schuster}}, \bibinfo {author} {\bibfnamefont {J.}~\bibnamefont {Simon}}, \bibinfo {author} {\bibfnamefont {O.}~\bibnamefont {Zilberberg}},\ and\ \bibinfo {author} {\bibfnamefont {I.}~\bibnamefont {Carusotto}},\ }\bibfield  {title} {\bibinfo {title} {Topological photonics},\ }\href {https://doi.org/10.1103/RevModPhys.91.015006} {\bibfield  {journal} {\bibinfo  {journal} {Rev. Mod. Phys.}\ }\textbf {\bibinfo {volume} {91}},\ \bibinfo {pages} {015006} (\bibinfo {year} {2019})}\BibitemShut {NoStop}%
\bibitem [{\citenamefont {Haldane}\ and\ \citenamefont {Raghu}(2008)}]{Haldane2008}%
  \BibitemOpen
  \bibfield  {author} {\bibinfo {author} {\bibfnamefont {F.~D.~M.}\ \bibnamefont {Haldane}}\ and\ \bibinfo {author} {\bibfnamefont {S.}~\bibnamefont {Raghu}},\ }\bibfield  {title} {\bibinfo {title} {Possible realization of directional optical waveguides in photonic crystals with broken time-reversal symmetry},\ }\href {https://doi.org/10.1103/PhysRevLett.100.013904} {\bibfield  {journal} {\bibinfo  {journal} {Phys. Rev. Lett.}\ }\textbf {\bibinfo {volume} {100}},\ \bibinfo {pages} {013904} (\bibinfo {year} {2008})}\BibitemShut {NoStop}%
\bibitem [{\citenamefont {Raghu}\ and\ \citenamefont {Haldane}(2008)}]{Raghu2008}%
  \BibitemOpen
  \bibfield  {author} {\bibinfo {author} {\bibfnamefont {S.}~\bibnamefont {Raghu}}\ and\ \bibinfo {author} {\bibfnamefont {F.~D.~M.}\ \bibnamefont {Haldane}},\ }\bibfield  {title} {\bibinfo {title} {Analogs of quantum-hall-effect edge states in photonic crystals},\ }\href {https://doi.org/10.1103/PhysRevA.78.033834} {\bibfield  {journal} {\bibinfo  {journal} {Phys. Rev. A}\ }\textbf {\bibinfo {volume} {78}},\ \bibinfo {pages} {033834} (\bibinfo {year} {2008})}\BibitemShut {NoStop}%
\bibitem [{\citenamefont {Wang}\ \emph {et~al.}(2008)\citenamefont {Wang}, \citenamefont {Chong}, \citenamefont {Joannopoulos},\ and\ \citenamefont {Solja\ifmmode \check{c}\else \v{c}\fi{}i\ifmmode~\acute{c}\else \'{c}\fi{}}}]{Wang2008}%
  \BibitemOpen
  \bibfield  {author} {\bibinfo {author} {\bibfnamefont {Z.}~\bibnamefont {Wang}}, \bibinfo {author} {\bibfnamefont {Y.~D.}\ \bibnamefont {Chong}}, \bibinfo {author} {\bibfnamefont {J.~D.}\ \bibnamefont {Joannopoulos}},\ and\ \bibinfo {author} {\bibfnamefont {M.}~\bibnamefont {Solja\ifmmode \check{c}\else \v{c}\fi{}i\ifmmode~\acute{c}\else \'{c}\fi{}}},\ }\bibfield  {title} {\bibinfo {title} {Reflection-free one-way edge modes in a gyromagnetic photonic crystal},\ }\href {https://doi.org/10.1103/PhysRevLett.100.013905} {\bibfield  {journal} {\bibinfo  {journal} {Phys. Rev. Lett.}\ }\textbf {\bibinfo {volume} {100}},\ \bibinfo {pages} {013905} (\bibinfo {year} {2008})}\BibitemShut {NoStop}%
\bibitem [{\citenamefont {Wang}\ \emph {et~al.}(2009)\citenamefont {Wang}, \citenamefont {Chong}, \citenamefont {Joannopoulos},\ and\ \citenamefont {Soljačić}}]{Wang2009}%
  \BibitemOpen
  \bibfield  {author} {\bibinfo {author} {\bibfnamefont {Z.}~\bibnamefont {Wang}}, \bibinfo {author} {\bibfnamefont {Y.}~\bibnamefont {Chong}}, \bibinfo {author} {\bibfnamefont {J.~D.}\ \bibnamefont {Joannopoulos}},\ and\ \bibinfo {author} {\bibfnamefont {M.}~\bibnamefont {Soljačić}},\ }\bibfield  {title} {\bibinfo {title} {Observation of unidirectional backscattering-immune topological electromagnetic states},\ }\href {https://doi.org/10.1038/nature08293} {\bibfield  {journal} {\bibinfo  {journal} {Nature}\ }\textbf {\bibinfo {volume} {461}},\ \bibinfo {pages} {772} (\bibinfo {year} {2009})}\BibitemShut {NoStop}%
\bibitem [{\citenamefont {Kim}\ \emph {et~al.}(2022)\citenamefont {Kim}, \citenamefont {Wang}, \citenamefont {Yang}, \citenamefont {Teo}, \citenamefont {Rho},\ and\ \citenamefont {Zhang}}]{Baile2022}%
  \BibitemOpen
  \bibfield  {author} {\bibinfo {author} {\bibfnamefont {M.}~\bibnamefont {Kim}}, \bibinfo {author} {\bibfnamefont {Z.}~\bibnamefont {Wang}}, \bibinfo {author} {\bibfnamefont {Y.}~\bibnamefont {Yang}}, \bibinfo {author} {\bibfnamefont {H.~T.}\ \bibnamefont {Teo}}, \bibinfo {author} {\bibfnamefont {J.}~\bibnamefont {Rho}},\ and\ \bibinfo {author} {\bibfnamefont {B.}~\bibnamefont {Zhang}},\ }\bibfield  {title} {\bibinfo {title} {Three-dimensional photonic topological insulator without spin–orbit coupling},\ }\href {https://doi.org/10.1038/s41467-022-30909-0} {\bibfield  {journal} {\bibinfo  {journal} {Nature Communications}\ }\textbf {\bibinfo {volume} {13}},\ \bibinfo {pages} {3499} (\bibinfo {year} {2022})}\BibitemShut {NoStop}%
\bibitem [{\citenamefont {Proctor}\ \emph {et~al.}(2020)\citenamefont {Proctor}, \citenamefont {Huidobro}, \citenamefont {Bradlyn}, \citenamefont {de~Paz}, \citenamefont {Vergniory}, \citenamefont {Bercioux},\ and\ \citenamefont {Garc\'{\i}a-Etxarri}}]{Etxarri2020}%
  \BibitemOpen
  \bibfield  {author} {\bibinfo {author} {\bibfnamefont {M.}~\bibnamefont {Proctor}}, \bibinfo {author} {\bibfnamefont {P.~A.}\ \bibnamefont {Huidobro}}, \bibinfo {author} {\bibfnamefont {B.}~\bibnamefont {Bradlyn}}, \bibinfo {author} {\bibfnamefont {M.~B.}\ \bibnamefont {de~Paz}}, \bibinfo {author} {\bibfnamefont {M.~G.}\ \bibnamefont {Vergniory}}, \bibinfo {author} {\bibfnamefont {D.}~\bibnamefont {Bercioux}},\ and\ \bibinfo {author} {\bibfnamefont {A.}~\bibnamefont {Garc\'{\i}a-Etxarri}},\ }\bibfield  {title} {\bibinfo {title} {Robustness of topological corner modes in photonic crystals},\ }\href {https://doi.org/10.1103/PhysRevResearch.2.042038} {\bibfield  {journal} {\bibinfo  {journal} {Phys. Rev. Res.}\ }\textbf {\bibinfo {volume} {2}},\ \bibinfo {pages} {042038} (\bibinfo {year} {2020})}\BibitemShut {NoStop}%
\bibitem [{\citenamefont {Bello}\ \emph {et~al.}(2019)\citenamefont {Bello}, \citenamefont {Platero}, \citenamefont {Cirac},\ and\ \citenamefont {González-Tudela}}]{Bello2019}%
  \BibitemOpen
  \bibfield  {author} {\bibinfo {author} {\bibfnamefont {M.}~\bibnamefont {Bello}}, \bibinfo {author} {\bibfnamefont {G.}~\bibnamefont {Platero}}, \bibinfo {author} {\bibfnamefont {J.~I.}\ \bibnamefont {Cirac}},\ and\ \bibinfo {author} {\bibfnamefont {A.}~\bibnamefont {González-Tudela}},\ }\bibfield  {title} {\bibinfo {title} {Unconventional quantum optics in topological waveguide qed},\ }\href {https://doi.org/10.1126/sciadv.aaw0297} {\bibfield  {journal} {\bibinfo  {journal} {Science Advances}\ }\textbf {\bibinfo {volume} {5}},\ \bibinfo {pages} {eaaw0297} (\bibinfo {year} {2019})}\BibitemShut {NoStop}%
\bibitem [{\citenamefont {Ke}\ \emph {et~al.}(2016)\citenamefont {Ke}, \citenamefont {Qin}, \citenamefont {Mei}, \citenamefont {Zhong}, \citenamefont {Kivshar},\ and\ \citenamefont {Lee}}]{Ke2016}%
  \BibitemOpen
  \bibfield  {author} {\bibinfo {author} {\bibfnamefont {Y.}~\bibnamefont {Ke}}, \bibinfo {author} {\bibfnamefont {X.}~\bibnamefont {Qin}}, \bibinfo {author} {\bibfnamefont {F.}~\bibnamefont {Mei}}, \bibinfo {author} {\bibfnamefont {H.}~\bibnamefont {Zhong}}, \bibinfo {author} {\bibfnamefont {Y.~S.}\ \bibnamefont {Kivshar}},\ and\ \bibinfo {author} {\bibfnamefont {C.}~\bibnamefont {Lee}},\ }\bibfield  {title} {\bibinfo {title} {Topological phase transitions and thouless pumping of light in photonic waveguide arrays},\ }\href {https://doi.org/https://doi.org/10.1002/lpor.201600119} {\bibfield  {journal} {\bibinfo  {journal} {Laser \& Photonics Reviews}\ }\textbf {\bibinfo {volume} {10}},\ \bibinfo {pages} {995} (\bibinfo {year} {2016})}\BibitemShut {NoStop}%
\bibitem [{\citenamefont {Mann}\ and\ \citenamefont {Mariani}(2022)}]{Mann2022}%
  \BibitemOpen
  \bibfield  {author} {\bibinfo {author} {\bibfnamefont {C.-R.}\ \bibnamefont {Mann}}\ and\ \bibinfo {author} {\bibfnamefont {E.}~\bibnamefont {Mariani}},\ }\bibfield  {title} {\bibinfo {title} {Topological transitions in arrays of dipoles coupled to a cavity waveguide},\ }\href {https://doi.org/10.1103/PhysRevResearch.4.013078} {\bibfield  {journal} {\bibinfo  {journal} {Phys. Rev. Res.}\ }\textbf {\bibinfo {volume} {4}},\ \bibinfo {pages} {013078} (\bibinfo {year} {2022})}\BibitemShut {NoStop}%
\bibitem [{\citenamefont {Nie}\ and\ \citenamefont {Liu}(2020)}]{Nie2020}%
  \BibitemOpen
  \bibfield  {author} {\bibinfo {author} {\bibfnamefont {W.}~\bibnamefont {Nie}}\ and\ \bibinfo {author} {\bibfnamefont {Y.-x.}\ \bibnamefont {Liu}},\ }\bibfield  {title} {\bibinfo {title} {Bandgap-assisted quantum control of topological edge states in a cavity},\ }\href {https://doi.org/10.1103/PhysRevResearch.2.012076} {\bibfield  {journal} {\bibinfo  {journal} {Phys. Rev. Res.}\ }\textbf {\bibinfo {volume} {2}},\ \bibinfo {pages} {012076} (\bibinfo {year} {2020})}\BibitemShut {NoStop}%
\bibitem [{\citenamefont {Wang}\ \emph {et~al.}(2016)\citenamefont {Wang}, \citenamefont {Yang}, \citenamefont {Hu}, \citenamefont {Xue},\ and\ \citenamefont {Wu}}]{Wang2016}%
  \BibitemOpen
  \bibfield  {author} {\bibinfo {author} {\bibfnamefont {Y.-P.}\ \bibnamefont {Wang}}, \bibinfo {author} {\bibfnamefont {W.-L.}\ \bibnamefont {Yang}}, \bibinfo {author} {\bibfnamefont {Y.}~\bibnamefont {Hu}}, \bibinfo {author} {\bibfnamefont {Z.-Y.}\ \bibnamefont {Xue}},\ and\ \bibinfo {author} {\bibfnamefont {Y.}~\bibnamefont {Wu}},\ }\bibfield  {title} {\bibinfo {title} {Detecting topological phases of microwave photons in a circuit quantum electrodynamics lattice},\ }\href {https://doi.org/10.1038/npjqi.2016.15} {\bibfield  {journal} {\bibinfo  {journal} {npj Quantum Information}\ }\textbf {\bibinfo {volume} {2}},\ \bibinfo {pages} {16015} (\bibinfo {year} {2016})}\BibitemShut {NoStop}%
\bibitem [{\citenamefont {St-Jean}\ \emph {et~al.}(2017)\citenamefont {St-Jean}, \citenamefont {Goblot}, \citenamefont {Galopin}, \citenamefont {Lemaître}, \citenamefont {Ozawa}, \citenamefont {Le~Gratiet}, \citenamefont {Sagnes}, \citenamefont {Bloch},\ and\ \citenamefont {Amo}}]{StJean2017}%
  \BibitemOpen
  \bibfield  {author} {\bibinfo {author} {\bibfnamefont {P.}~\bibnamefont {St-Jean}}, \bibinfo {author} {\bibfnamefont {V.}~\bibnamefont {Goblot}}, \bibinfo {author} {\bibfnamefont {E.}~\bibnamefont {Galopin}}, \bibinfo {author} {\bibfnamefont {A.}~\bibnamefont {Lemaître}}, \bibinfo {author} {\bibfnamefont {T.}~\bibnamefont {Ozawa}}, \bibinfo {author} {\bibfnamefont {L.}~\bibnamefont {Le~Gratiet}}, \bibinfo {author} {\bibfnamefont {I.}~\bibnamefont {Sagnes}}, \bibinfo {author} {\bibfnamefont {J.}~\bibnamefont {Bloch}},\ and\ \bibinfo {author} {\bibfnamefont {A.}~\bibnamefont {Amo}},\ }\bibfield  {title} {\bibinfo {title} {Lasing in topological edge states of a one-dimensional lattice},\ }\href {https://doi.org/10.1038/s41566-017-0006-2} {\bibfield  {journal} {\bibinfo  {journal} {Nature Photonics}\ }\textbf {\bibinfo {volume} {11}},\ \bibinfo {pages} {651} (\bibinfo {year} {2017})}\BibitemShut {NoStop}%
\bibitem [{\citenamefont {Bandres}\ \emph {et~al.}(2018)\citenamefont {Bandres}, \citenamefont {Wittek}, \citenamefont {Harari}, \citenamefont {Parto}, \citenamefont {Ren}, \citenamefont {Segev}, \citenamefont {Christodoulides},\ and\ \citenamefont {Khajavikhan}}]{Bandres2018}%
  \BibitemOpen
  \bibfield  {author} {\bibinfo {author} {\bibfnamefont {M.~A.}\ \bibnamefont {Bandres}}, \bibinfo {author} {\bibfnamefont {S.}~\bibnamefont {Wittek}}, \bibinfo {author} {\bibfnamefont {G.}~\bibnamefont {Harari}}, \bibinfo {author} {\bibfnamefont {M.}~\bibnamefont {Parto}}, \bibinfo {author} {\bibfnamefont {J.}~\bibnamefont {Ren}}, \bibinfo {author} {\bibfnamefont {M.}~\bibnamefont {Segev}}, \bibinfo {author} {\bibfnamefont {D.~N.}\ \bibnamefont {Christodoulides}},\ and\ \bibinfo {author} {\bibfnamefont {M.}~\bibnamefont {Khajavikhan}},\ }\bibfield  {title} {\bibinfo {title} {Topological insulator laser: Experiments},\ }\href {https://doi.org/10.1126/science.aar4005} {\bibfield  {journal} {\bibinfo  {journal} {Science}\ }\textbf {\bibinfo {volume} {359}},\ \bibinfo {pages} {eaar4005} (\bibinfo {year} {2018})}\BibitemShut {NoStop}%
\bibitem [{\citenamefont {Kitaev}(2003)}]{Kitaev2003}%
  \BibitemOpen
  \bibfield  {author} {\bibinfo {author} {\bibfnamefont {A.}~\bibnamefont {Kitaev}},\ }\bibfield  {title} {\bibinfo {title} {Fault-tolerant quantum computation by anyons},\ }\href {https://doi.org/https://doi.org/10.1016/S0003-4916(02)00018-0} {\bibfield  {journal} {\bibinfo  {journal} {Annals of Physics}\ }\textbf {\bibinfo {volume} {303}},\ \bibinfo {pages} {2} (\bibinfo {year} {2003})}\BibitemShut {NoStop}%
\bibitem [{\citenamefont {Nayak}\ \emph {et~al.}(2008)\citenamefont {Nayak}, \citenamefont {Simon}, \citenamefont {Stern}, \citenamefont {Freedman},\ and\ \citenamefont {Das~Sarma}}]{Nayak2008}%
  \BibitemOpen
  \bibfield  {author} {\bibinfo {author} {\bibfnamefont {C.}~\bibnamefont {Nayak}}, \bibinfo {author} {\bibfnamefont {S.~H.}\ \bibnamefont {Simon}}, \bibinfo {author} {\bibfnamefont {A.}~\bibnamefont {Stern}}, \bibinfo {author} {\bibfnamefont {M.}~\bibnamefont {Freedman}},\ and\ \bibinfo {author} {\bibfnamefont {S.}~\bibnamefont {Das~Sarma}},\ }\bibfield  {title} {\bibinfo {title} {Non-abelian anyons and topological quantum computation},\ }\href {https://doi.org/10.1103/RevModPhys.80.1083} {\bibfield  {journal} {\bibinfo  {journal} {Rev. Mod. Phys.}\ }\textbf {\bibinfo {volume} {80}},\ \bibinfo {pages} {1083} (\bibinfo {year} {2008})}\BibitemShut {NoStop}%
\bibitem [{\citenamefont {Field}\ and\ \citenamefont {Simula}(2018)}]{Field2018}%
  \BibitemOpen
  \bibfield  {author} {\bibinfo {author} {\bibfnamefont {B.}~\bibnamefont {Field}}\ and\ \bibinfo {author} {\bibfnamefont {T.}~\bibnamefont {Simula}},\ }\bibfield  {title} {\bibinfo {title} {Introduction to topological quantum computation with non-abelian anyons},\ }\href {https://doi.org/10.1088/2058-9565/aacad2} {\bibfield  {journal} {\bibinfo  {journal} {Quantum Science and Technology}\ }\textbf {\bibinfo {volume} {3}},\ \bibinfo {pages} {045004} (\bibinfo {year} {2018})}\BibitemShut {NoStop}%
\bibitem [{\citenamefont {D'Angelis}\ \emph {et~al.}(2020)\citenamefont {D'Angelis}, \citenamefont {Pinheiro}, \citenamefont {Gu\'ery-Odelin}, \citenamefont {Longhi},\ and\ \citenamefont {Impens}}]{Angelis2020}%
  \BibitemOpen
  \bibfield  {author} {\bibinfo {author} {\bibfnamefont {F.~M.}\ \bibnamefont {D'Angelis}}, \bibinfo {author} {\bibfnamefont {F.~A.}\ \bibnamefont {Pinheiro}}, \bibinfo {author} {\bibfnamefont {D.}~\bibnamefont {Gu\'ery-Odelin}}, \bibinfo {author} {\bibfnamefont {S.}~\bibnamefont {Longhi}},\ and\ \bibinfo {author} {\bibfnamefont {F.}~\bibnamefont {Impens}},\ }\bibfield  {title} {\bibinfo {title} {Fast and robust quantum state transfer in a topological su-schrieffer-heeger chain with next-to-nearest-neighbor interactions},\ }\href {https://doi.org/10.1103/PhysRevResearch.2.033475} {\bibfield  {journal} {\bibinfo  {journal} {Phys. Rev. Res.}\ }\textbf {\bibinfo {volume} {2}},\ \bibinfo {pages} {033475} (\bibinfo {year} {2020})}\BibitemShut {NoStop}%
\bibitem [{\citenamefont {Lang}\ and\ \citenamefont {Büchler}(2017)}]{Lang2017}%
  \BibitemOpen
  \bibfield  {author} {\bibinfo {author} {\bibfnamefont {N.}~\bibnamefont {Lang}}\ and\ \bibinfo {author} {\bibfnamefont {H.~P.}\ \bibnamefont {Büchler}},\ }\bibfield  {title} {\bibinfo {title} {Topological networks for quantum communication between distant qubits},\ }\href {https://doi.org/10.1038/s41534-017-0047-x} {\bibfield  {journal} {\bibinfo  {journal} {npj Quantum Inf.}\ }\textbf {\bibinfo {volume} {3}},\ \bibinfo {pages} {47} (\bibinfo {year} {2017})}\BibitemShut {NoStop}%
\bibitem [{\citenamefont {Chang}\ \emph {et~al.}(2018)\citenamefont {Chang}, \citenamefont {Douglas}, \citenamefont {Gonz\'alez-Tudela}, \citenamefont {Hung},\ and\ \citenamefont {Kimble}}]{Chang2018}%
  \BibitemOpen
  \bibfield  {author} {\bibinfo {author} {\bibfnamefont {D.~E.}\ \bibnamefont {Chang}}, \bibinfo {author} {\bibfnamefont {J.~S.}\ \bibnamefont {Douglas}}, \bibinfo {author} {\bibfnamefont {A.}~\bibnamefont {Gonz\'alez-Tudela}}, \bibinfo {author} {\bibfnamefont {C.-L.}\ \bibnamefont {Hung}},\ and\ \bibinfo {author} {\bibfnamefont {H.~J.}\ \bibnamefont {Kimble}},\ }\bibfield  {title} {\bibinfo {title} {Colloquium: Quantum matter built from nanoscopic lattices of atoms and photons},\ }\href {https://doi.org/10.1103/RevModPhys.90.031002} {\bibfield  {journal} {\bibinfo  {journal} {Rev. Mod. Phys.}\ }\textbf {\bibinfo {volume} {90}},\ \bibinfo {pages} {031002} (\bibinfo {year} {2018})}\BibitemShut {NoStop}%
\bibitem [{\citenamefont {Gross}\ and\ \citenamefont {Bloch}(2017)}]{Bloch2017}%
  \BibitemOpen
  \bibfield  {author} {\bibinfo {author} {\bibfnamefont {C.}~\bibnamefont {Gross}}\ and\ \bibinfo {author} {\bibfnamefont {I.}~\bibnamefont {Bloch}},\ }\bibfield  {title} {\bibinfo {title} {Quantum simulations with ultracold atoms in optical lattices},\ }\href {https://doi.org/10.1126/science.aal3837} {\bibfield  {journal} {\bibinfo  {journal} {Science}\ }\textbf {\bibinfo {volume} {357}},\ \bibinfo {pages} {995} (\bibinfo {year} {2017})}\BibitemShut {NoStop}%
\bibitem [{\citenamefont {Cooper}\ \emph {et~al.}(2019)\citenamefont {Cooper}, \citenamefont {Dalibard},\ and\ \citenamefont {Spielman}}]{Cooper2019}%
  \BibitemOpen
  \bibfield  {author} {\bibinfo {author} {\bibfnamefont {N.~R.}\ \bibnamefont {Cooper}}, \bibinfo {author} {\bibfnamefont {J.}~\bibnamefont {Dalibard}},\ and\ \bibinfo {author} {\bibfnamefont {I.~B.}\ \bibnamefont {Spielman}},\ }\bibfield  {title} {\bibinfo {title} {Topological bands for ultracold atoms},\ }\href {https://doi.org/10.1103/RevModPhys.91.015005} {\bibfield  {journal} {\bibinfo  {journal} {Rev. Mod. Phys.}\ }\textbf {\bibinfo {volume} {91}},\ \bibinfo {pages} {015005} (\bibinfo {year} {2019})}\BibitemShut {NoStop}%
\bibitem [{\citenamefont {de~Léséleuc}\ \emph {et~al.}(2019)\citenamefont {de~Léséleuc}, \citenamefont {Lienhard}, \citenamefont {Scholl}, \citenamefont {Barredo}, \citenamefont {Weber}, \citenamefont {Lang}, \citenamefont {Büchler}, \citenamefont {Lahaye},\ and\ \citenamefont {Browaeys}}]{Broaweys2019}%
  \BibitemOpen
  \bibfield  {author} {\bibinfo {author} {\bibfnamefont {S.}~\bibnamefont {de~Léséleuc}}, \bibinfo {author} {\bibfnamefont {V.}~\bibnamefont {Lienhard}}, \bibinfo {author} {\bibfnamefont {P.}~\bibnamefont {Scholl}}, \bibinfo {author} {\bibfnamefont {D.}~\bibnamefont {Barredo}}, \bibinfo {author} {\bibfnamefont {S.}~\bibnamefont {Weber}}, \bibinfo {author} {\bibfnamefont {N.}~\bibnamefont {Lang}}, \bibinfo {author} {\bibfnamefont {H.~P.}\ \bibnamefont {Büchler}}, \bibinfo {author} {\bibfnamefont {T.}~\bibnamefont {Lahaye}},\ and\ \bibinfo {author} {\bibfnamefont {A.}~\bibnamefont {Browaeys}},\ }\bibfield  {title} {\bibinfo {title} {Observation of a symmetry-protected topological phase of interacting bosons with rydberg atoms},\ }\href {https://doi.org/10.1126/science.aav9105} {\bibfield  {journal} {\bibinfo  {journal} {Science}\ }\textbf {\bibinfo {volume} {365}},\ \bibinfo {pages} {775} (\bibinfo {year} {2019})}\BibitemShut {NoStop}%
\bibitem [{\citenamefont {Blackmore}\ \emph {et~al.}(2018)\citenamefont {Blackmore}, \citenamefont {Caldwell}, \citenamefont {Gregory}, \citenamefont {Bridge}, \citenamefont {Sawant}, \citenamefont {Aldegunde}, \citenamefont {Mur-Petit}, \citenamefont {Jaksch}, \citenamefont {Hutson}, \citenamefont {Sauer}, \citenamefont {Tarbutt},\ and\ \citenamefont {Cornish}}]{Blackmore2019}%
  \BibitemOpen
  \bibfield  {author} {\bibinfo {author} {\bibfnamefont {J.~A.}\ \bibnamefont {Blackmore}}, \bibinfo {author} {\bibfnamefont {L.}~\bibnamefont {Caldwell}}, \bibinfo {author} {\bibfnamefont {P.~D.}\ \bibnamefont {Gregory}}, \bibinfo {author} {\bibfnamefont {E.~M.}\ \bibnamefont {Bridge}}, \bibinfo {author} {\bibfnamefont {R.}~\bibnamefont {Sawant}}, \bibinfo {author} {\bibfnamefont {J.}~\bibnamefont {Aldegunde}}, \bibinfo {author} {\bibfnamefont {J.}~\bibnamefont {Mur-Petit}}, \bibinfo {author} {\bibfnamefont {D.}~\bibnamefont {Jaksch}}, \bibinfo {author} {\bibfnamefont {J.~M.}\ \bibnamefont {Hutson}}, \bibinfo {author} {\bibfnamefont {B.~E.}\ \bibnamefont {Sauer}}, \bibinfo {author} {\bibfnamefont {M.~R.}\ \bibnamefont {Tarbutt}},\ and\ \bibinfo {author} {\bibfnamefont {S.~L.}\ \bibnamefont {Cornish}},\ }\bibfield  {title} {\bibinfo {title} {Ultracold molecules for quantum simulation: rotational coherences in caf and rbcs},\ }\href {https://doi.org/10.1088/2058-9565/aaee35} {\bibfield  {journal} {\bibinfo
  {journal} {Quantum Science and Technology}\ }\textbf {\bibinfo {volume} {4}},\ \bibinfo {pages} {014010} (\bibinfo {year} {2018})}\BibitemShut {NoStop}%
\bibitem [{\citenamefont {Bao}\ \emph {et~al.}(2023)\citenamefont {Bao}, \citenamefont {Yu}, \citenamefont {Anderegg}, \citenamefont {Chae}, \citenamefont {Ketterle}, \citenamefont {Ni},\ and\ \citenamefont {Doyle}}]{Yicheng2023}%
  \BibitemOpen
  \bibfield  {author} {\bibinfo {author} {\bibfnamefont {Y.}~\bibnamefont {Bao}}, \bibinfo {author} {\bibfnamefont {S.~S.}\ \bibnamefont {Yu}}, \bibinfo {author} {\bibfnamefont {L.}~\bibnamefont {Anderegg}}, \bibinfo {author} {\bibfnamefont {E.}~\bibnamefont {Chae}}, \bibinfo {author} {\bibfnamefont {W.}~\bibnamefont {Ketterle}}, \bibinfo {author} {\bibfnamefont {K.-K.}\ \bibnamefont {Ni}},\ and\ \bibinfo {author} {\bibfnamefont {J.~M.}\ \bibnamefont {Doyle}},\ }\bibfield  {title} {\bibinfo {title} {Dipolar spin-exchange and entanglement between molecules in an optical tweezer array},\ }\href {https://doi.org/10.1126/science.adf8999} {\bibfield  {journal} {\bibinfo  {journal} {Science}\ }\textbf {\bibinfo {volume} {382}},\ \bibinfo {pages} {1138} (\bibinfo {year} {2023})}\BibitemShut {NoStop}%
\bibitem [{\citenamefont {Holland}\ \emph {et~al.}(2023)\citenamefont {Holland}, \citenamefont {Lu},\ and\ \citenamefont {Cheuk}}]{Connor2023}%
  \BibitemOpen
  \bibfield  {author} {\bibinfo {author} {\bibfnamefont {C.~M.}\ \bibnamefont {Holland}}, \bibinfo {author} {\bibfnamefont {Y.}~\bibnamefont {Lu}},\ and\ \bibinfo {author} {\bibfnamefont {L.~W.}\ \bibnamefont {Cheuk}},\ }\bibfield  {title} {\bibinfo {title} {On-demand entanglement of molecules in a reconfigurable optical tweezer array},\ }\href {https://doi.org/10.1126/science.adf4272} {\bibfield  {journal} {\bibinfo  {journal} {Science}\ }\textbf {\bibinfo {volume} {382}},\ \bibinfo {pages} {1143} (\bibinfo {year} {2023})}\BibitemShut {NoStop}%
\bibitem [{\citenamefont {Slobozhanyuk}\ \emph {et~al.}(2015)\citenamefont {Slobozhanyuk}, \citenamefont {Poddubny}, \citenamefont {Miroshnichenko}, \citenamefont {Belov},\ and\ \citenamefont {Kivshar}}]{Kivshar2015}%
  \BibitemOpen
  \bibfield  {author} {\bibinfo {author} {\bibfnamefont {A.~P.}\ \bibnamefont {Slobozhanyuk}}, \bibinfo {author} {\bibfnamefont {A.~N.}\ \bibnamefont {Poddubny}}, \bibinfo {author} {\bibfnamefont {A.~E.}\ \bibnamefont {Miroshnichenko}}, \bibinfo {author} {\bibfnamefont {P.~A.}\ \bibnamefont {Belov}},\ and\ \bibinfo {author} {\bibfnamefont {Y.~S.}\ \bibnamefont {Kivshar}},\ }\bibfield  {title} {\bibinfo {title} {Subwavelength topological edge states in optically resonant dielectric structures},\ }\href {https://doi.org/10.1103/PhysRevLett.114.123901} {\bibfield  {journal} {\bibinfo  {journal} {Phys. Rev. Lett.}\ }\textbf {\bibinfo {volume} {114}},\ \bibinfo {pages} {123901} (\bibinfo {year} {2015})}\BibitemShut {NoStop}%
\bibitem [{\citenamefont {Wang}\ and\ \citenamefont {Zhao}(2018)}]{Wang2018}%
  \BibitemOpen
  \bibfield  {author} {\bibinfo {author} {\bibfnamefont {B.~X.}\ \bibnamefont {Wang}}\ and\ \bibinfo {author} {\bibfnamefont {C.~Y.}\ \bibnamefont {Zhao}},\ }\bibfield  {title} {\bibinfo {title} {Topological phonon polaritons in one-dimensional non-hermitian silicon carbide nanoparticle chains},\ }\href {https://doi.org/10.1103/PhysRevB.98.165435} {\bibfield  {journal} {\bibinfo  {journal} {Phys. Rev. B}\ }\textbf {\bibinfo {volume} {98}},\ \bibinfo {pages} {165435} (\bibinfo {year} {2018})}\BibitemShut {NoStop}%
\bibitem [{\citenamefont {Downing}\ and\ \citenamefont {Weick}(2017)}]{Downing2017}%
  \BibitemOpen
  \bibfield  {author} {\bibinfo {author} {\bibfnamefont {C.~A.}\ \bibnamefont {Downing}}\ and\ \bibinfo {author} {\bibfnamefont {G.}~\bibnamefont {Weick}},\ }\bibfield  {title} {\bibinfo {title} {Topological collective plasmons in bipartite chains of metallic nanoparticles},\ }\href {https://doi.org/10.1103/PhysRevB.95.125426} {\bibfield  {journal} {\bibinfo  {journal} {Phys. Rev. B}\ }\textbf {\bibinfo {volume} {95}},\ \bibinfo {pages} {125426} (\bibinfo {year} {2017})}\BibitemShut {NoStop}%
\bibitem [{\citenamefont {Bello}\ and\ \citenamefont {Cirac}(2023)}]{Cirac2023}%
  \BibitemOpen
  \bibfield  {author} {\bibinfo {author} {\bibfnamefont {M.}~\bibnamefont {Bello}}\ and\ \bibinfo {author} {\bibfnamefont {J.~I.}\ \bibnamefont {Cirac}},\ }\bibfield  {title} {\bibinfo {title} {Topological effects in two-dimensional quantum emitter systems},\ }\href {https://doi.org/10.1103/PhysRevB.107.054301} {\bibfield  {journal} {\bibinfo  {journal} {Phys. Rev. B}\ }\textbf {\bibinfo {volume} {107}},\ \bibinfo {pages} {054301} (\bibinfo {year} {2023})}\BibitemShut {NoStop}%
\bibitem [{\citenamefont {Gong}\ \emph {et~al.}(2016)\citenamefont {Gong}, \citenamefont {Maghrebi}, \citenamefont {Hu}, \citenamefont {Wall}, \citenamefont {Foss-Feig},\ and\ \citenamefont {Gorshkov}}]{Gong2016}%
  \BibitemOpen
  \bibfield  {author} {\bibinfo {author} {\bibfnamefont {Z.-X.}\ \bibnamefont {Gong}}, \bibinfo {author} {\bibfnamefont {M.~F.}\ \bibnamefont {Maghrebi}}, \bibinfo {author} {\bibfnamefont {A.}~\bibnamefont {Hu}}, \bibinfo {author} {\bibfnamefont {M.~L.}\ \bibnamefont {Wall}}, \bibinfo {author} {\bibfnamefont {M.}~\bibnamefont {Foss-Feig}},\ and\ \bibinfo {author} {\bibfnamefont {A.~V.}\ \bibnamefont {Gorshkov}},\ }\bibfield  {title} {\bibinfo {title} {Topological phases with long-range interactions},\ }\href {https://doi.org/10.1103/PhysRevB.93.041102} {\bibfield  {journal} {\bibinfo  {journal} {Phys. Rev. B}\ }\textbf {\bibinfo {volume} {93}},\ \bibinfo {pages} {041102} (\bibinfo {year} {2016})}\BibitemShut {NoStop}%
\bibitem [{\citenamefont {Lepori}\ and\ \citenamefont {Dell’Anna}(2017)}]{Lepori2017}%
  \BibitemOpen
  \bibfield  {author} {\bibinfo {author} {\bibfnamefont {L.}~\bibnamefont {Lepori}}\ and\ \bibinfo {author} {\bibfnamefont {L.}~\bibnamefont {Dell’Anna}},\ }\bibfield  {title} {\bibinfo {title} {Long-range topological insulators and weakened bulk-boundary correspondence},\ }\href {https://doi.org/10.1088/1367-2630/aa84d0} {\bibfield  {journal} {\bibinfo  {journal} {New Journal of Physics}\ }\textbf {\bibinfo {volume} {19}},\ \bibinfo {pages} {103030} (\bibinfo {year} {2017})}\BibitemShut {NoStop}%
\bibitem [{\citenamefont {Li}\ \emph {et~al.}(2014)\citenamefont {Li}, \citenamefont {Xu},\ and\ \citenamefont {Chen}}]{LiChen2014}%
  \BibitemOpen
  \bibfield  {author} {\bibinfo {author} {\bibfnamefont {L.}~\bibnamefont {Li}}, \bibinfo {author} {\bibfnamefont {Z.}~\bibnamefont {Xu}},\ and\ \bibinfo {author} {\bibfnamefont {S.}~\bibnamefont {Chen}},\ }\bibfield  {title} {\bibinfo {title} {Topological phases of generalized su-schrieffer-heeger models},\ }\href {https://doi.org/10.1103/PhysRevB.89.085111} {\bibfield  {journal} {\bibinfo  {journal} {Phys. Rev. B}\ }\textbf {\bibinfo {volume} {89}},\ \bibinfo {pages} {085111} (\bibinfo {year} {2014})}\BibitemShut {NoStop}%
\bibitem [{\citenamefont {P\'erez-Gonz\'alez}\ \emph {et~al.}(2019)\citenamefont {P\'erez-Gonz\'alez}, \citenamefont {Bello}, \citenamefont {G\'omez-Le\'on},\ and\ \citenamefont {Platero}}]{Platero2019}%
  \BibitemOpen
  \bibfield  {author} {\bibinfo {author} {\bibfnamefont {B.}~\bibnamefont {P\'erez-Gonz\'alez}}, \bibinfo {author} {\bibfnamefont {M.}~\bibnamefont {Bello}}, \bibinfo {author} {\bibfnamefont {A.}~\bibnamefont {G\'omez-Le\'on}},\ and\ \bibinfo {author} {\bibfnamefont {G.}~\bibnamefont {Platero}},\ }\bibfield  {title} {\bibinfo {title} {Interplay between long-range hopping and disorder in topological systems},\ }\href {https://doi.org/10.1103/PhysRevB.99.035146} {\bibfield  {journal} {\bibinfo  {journal} {Phys. Rev. B}\ }\textbf {\bibinfo {volume} {99}},\ \bibinfo {pages} {035146} (\bibinfo {year} {2019})}\BibitemShut {NoStop}%
\bibitem [{\citenamefont {Hsu}\ and\ \citenamefont {Chen}(2020)}]{HsuChen2020}%
  \BibitemOpen
  \bibfield  {author} {\bibinfo {author} {\bibfnamefont {H.-C.}\ \bibnamefont {Hsu}}\ and\ \bibinfo {author} {\bibfnamefont {T.-W.}\ \bibnamefont {Chen}},\ }\bibfield  {title} {\bibinfo {title} {Topological anderson insulating phases in the long-range su-schrieffer-heeger model},\ }\href {https://doi.org/10.1103/PhysRevB.102.205425} {\bibfield  {journal} {\bibinfo  {journal} {Phys. Rev. B}\ }\textbf {\bibinfo {volume} {102}},\ \bibinfo {pages} {205425} (\bibinfo {year} {2020})}\BibitemShut {NoStop}%
\bibitem [{\citenamefont {Appugliese}\ \emph {et~al.}(2022)\citenamefont {Appugliese}, \citenamefont {Enkner}, \citenamefont {Paravicini-Bagliani}, \citenamefont {Beck}, \citenamefont {Reichl}, \citenamefont {Wegscheider}, \citenamefont {Scalari}, \citenamefont {Ciuti},\ and\ \citenamefont {Faist}}]{Faist2022}%
  \BibitemOpen
  \bibfield  {author} {\bibinfo {author} {\bibfnamefont {F.}~\bibnamefont {Appugliese}}, \bibinfo {author} {\bibfnamefont {J.}~\bibnamefont {Enkner}}, \bibinfo {author} {\bibfnamefont {G.~L.}\ \bibnamefont {Paravicini-Bagliani}}, \bibinfo {author} {\bibfnamefont {M.}~\bibnamefont {Beck}}, \bibinfo {author} {\bibfnamefont {C.}~\bibnamefont {Reichl}}, \bibinfo {author} {\bibfnamefont {W.}~\bibnamefont {Wegscheider}}, \bibinfo {author} {\bibfnamefont {G.}~\bibnamefont {Scalari}}, \bibinfo {author} {\bibfnamefont {C.}~\bibnamefont {Ciuti}},\ and\ \bibinfo {author} {\bibfnamefont {J.}~\bibnamefont {Faist}},\ }\bibfield  {title} {\bibinfo {title} {Breakdown of topological protection by cavity vacuum fields in the integer quantum hall effect},\ }\href {https://doi.org/10.1126/science.abl5818} {\bibfield  {journal} {\bibinfo  {journal} {Science}\ }\textbf {\bibinfo {volume} {375}},\ \bibinfo {pages} {1030} (\bibinfo {year} {2022})}\BibitemShut {NoStop}%
\bibitem [{\citenamefont {Pocock}\ \emph {et~al.}(2019)\citenamefont {Pocock}, \citenamefont {Huidobro},\ and\ \citenamefont {Giannini}}]{Huidobro2019}%
  \BibitemOpen
  \bibfield  {author} {\bibinfo {author} {\bibfnamefont {S.~R.}\ \bibnamefont {Pocock}}, \bibinfo {author} {\bibfnamefont {P.~A.}\ \bibnamefont {Huidobro}},\ and\ \bibinfo {author} {\bibfnamefont {V.}~\bibnamefont {Giannini}},\ }\bibfield  {title} {\bibinfo {title} {Bulk-edge correspondence and long-range hopping in the topological plasmonic chain},\ }\href {https://doi.org/doi:10.1515/nanoph-2019-0033} {\bibfield  {journal} {\bibinfo  {journal} {Nanophotonics}\ }\textbf {\bibinfo {volume} {8}},\ \bibinfo {pages} {1337} (\bibinfo {year} {2019})}\BibitemShut {NoStop}%
\bibitem [{\citenamefont {Allard}\ and\ \citenamefont {Weick}(2023)}]{Allard2023}%
  \BibitemOpen
  \bibfield  {author} {\bibinfo {author} {\bibfnamefont {T.~F.}\ \bibnamefont {Allard}}\ and\ \bibinfo {author} {\bibfnamefont {G.}~\bibnamefont {Weick}},\ }\bibfield  {title} {\bibinfo {title} {Multiple polaritonic edge states in a su-schrieffer-heeger chain strongly coupled to a multimode cavity},\ }\href {https://doi.org/10.1103/PhysRevB.108.245417} {\bibfield  {journal} {\bibinfo  {journal} {Phys. Rev. B}\ }\textbf {\bibinfo {volume} {108}},\ \bibinfo {pages} {245417} (\bibinfo {year} {2023})}\BibitemShut {NoStop}%
\bibitem [{\citenamefont {Nie}\ \emph {et~al.}(2021{\natexlab{a}})\citenamefont {Nie}, \citenamefont {Antezza}, \citenamefont {Liu},\ and\ \citenamefont {Nori}}]{Nori2021}%
  \BibitemOpen
  \bibfield  {author} {\bibinfo {author} {\bibfnamefont {W.}~\bibnamefont {Nie}}, \bibinfo {author} {\bibfnamefont {M.}~\bibnamefont {Antezza}}, \bibinfo {author} {\bibfnamefont {Y.-x.}\ \bibnamefont {Liu}},\ and\ \bibinfo {author} {\bibfnamefont {F.}~\bibnamefont {Nori}},\ }\bibfield  {title} {\bibinfo {title} {Dissipative topological phase transition with strong system-environment coupling},\ }\href {https://doi.org/10.1103/PhysRevLett.127.250402} {\bibfield  {journal} {\bibinfo  {journal} {Phys. Rev. Lett.}\ }\textbf {\bibinfo {volume} {127}},\ \bibinfo {pages} {250402} (\bibinfo {year} {2021}{\natexlab{a}})}\BibitemShut {NoStop}%
\bibitem [{\citenamefont {Gong}\ \emph {et~al.}(2018)\citenamefont {Gong}, \citenamefont {Ashida}, \citenamefont {Kawabata}, \citenamefont {Takasan}, \citenamefont {Higashikawa},\ and\ \citenamefont {Ueda}}]{Gong2018}%
  \BibitemOpen
  \bibfield  {author} {\bibinfo {author} {\bibfnamefont {Z.}~\bibnamefont {Gong}}, \bibinfo {author} {\bibfnamefont {Y.}~\bibnamefont {Ashida}}, \bibinfo {author} {\bibfnamefont {K.}~\bibnamefont {Kawabata}}, \bibinfo {author} {\bibfnamefont {K.}~\bibnamefont {Takasan}}, \bibinfo {author} {\bibfnamefont {S.}~\bibnamefont {Higashikawa}},\ and\ \bibinfo {author} {\bibfnamefont {M.}~\bibnamefont {Ueda}},\ }\bibfield  {title} {\bibinfo {title} {Topological phases of non-hermitian systems},\ }\href {https://doi.org/10.1103/PhysRevX.8.031079} {\bibfield  {journal} {\bibinfo  {journal} {Phys. Rev. X}\ }\textbf {\bibinfo {volume} {8}},\ \bibinfo {pages} {031079} (\bibinfo {year} {2018})}\BibitemShut {NoStop}%
\bibitem [{\citenamefont {Svendsen}\ \emph {et~al.}(2024)\citenamefont {Svendsen}, \citenamefont {Cech}, \citenamefont {Schemmer},\ and\ \citenamefont {Olmos}}]{Svendsen2024}%
  \BibitemOpen
  \bibfield  {author} {\bibinfo {author} {\bibfnamefont {M.~B.~M.}\ \bibnamefont {Svendsen}}, \bibinfo {author} {\bibfnamefont {M.}~\bibnamefont {Cech}}, \bibinfo {author} {\bibfnamefont {M.}~\bibnamefont {Schemmer}},\ and\ \bibinfo {author} {\bibfnamefont {B.}~\bibnamefont {Olmos}},\ }\bibfield  {title} {\bibinfo {title} {Topological photon pumping in quantum optical systems},\ }\href {https://doi.org/10.22331/q-2024-10-02-1488} {\bibfield  {journal} {\bibinfo  {journal} {{Quantum}}\ }\textbf {\bibinfo {volume} {8}},\ \bibinfo {pages} {1488} (\bibinfo {year} {2024})}\BibitemShut {NoStop}%
\bibitem [{\citenamefont {Molesky}\ \emph {et~al.}(2018)\citenamefont {Molesky}, \citenamefont {Lin}, \citenamefont {Piggott}, \citenamefont {Jin}, \citenamefont {Vucković},\ and\ \citenamefont {Rodriguez}}]{Molesky2018}%
  \BibitemOpen
  \bibfield  {author} {\bibinfo {author} {\bibfnamefont {S.}~\bibnamefont {Molesky}}, \bibinfo {author} {\bibfnamefont {Z.}~\bibnamefont {Lin}}, \bibinfo {author} {\bibfnamefont {A.~Y.}\ \bibnamefont {Piggott}}, \bibinfo {author} {\bibfnamefont {W.}~\bibnamefont {Jin}}, \bibinfo {author} {\bibfnamefont {J.}~\bibnamefont {Vucković}},\ and\ \bibinfo {author} {\bibfnamefont {A.~W.}\ \bibnamefont {Rodriguez}},\ }\bibfield  {title} {\bibinfo {title} {Inverse design in nanophotonics},\ }\href {https://doi.org/10.1038/s41566-018-0246-9} {\bibfield  {journal} {\bibinfo  {journal} {Nature Photonics}\ }\textbf {\bibinfo {volume} {12}},\ \bibinfo {pages} {659} (\bibinfo {year} {2018})}\BibitemShut {NoStop}%
\bibitem [{\citenamefont {So}\ \emph {et~al.}(2020)\citenamefont {So}, \citenamefont {Badloe}, \citenamefont {Noh}, \citenamefont {Bravo-Abad},\ and\ \citenamefont {Rho}}]{BravoAbad2020}%
  \BibitemOpen
  \bibfield  {author} {\bibinfo {author} {\bibfnamefont {S.}~\bibnamefont {So}}, \bibinfo {author} {\bibfnamefont {T.}~\bibnamefont {Badloe}}, \bibinfo {author} {\bibfnamefont {J.}~\bibnamefont {Noh}}, \bibinfo {author} {\bibfnamefont {J.}~\bibnamefont {Bravo-Abad}},\ and\ \bibinfo {author} {\bibfnamefont {J.}~\bibnamefont {Rho}},\ }\bibfield  {title} {\bibinfo {title} {Deep learning enabled inverse design in nanophotonics},\ }\href {https://doi.org/doi:10.1515/nanoph-2019-0474} {\bibfield  {journal} {\bibinfo  {journal} {Nanophotonics}\ }\textbf {\bibinfo {volume} {9}},\ \bibinfo {pages} {1041} (\bibinfo {year} {2020})}\BibitemShut {NoStop}%
\bibitem [{\citenamefont {Mignuzzi}\ \emph {et~al.}(2019)\citenamefont {Mignuzzi}, \citenamefont {Vezzoli}, \citenamefont {Horsley}, \citenamefont {Barnes}, \citenamefont {Maier},\ and\ \citenamefont {Sapienza}}]{Mignuzzi2019}%
  \BibitemOpen
  \bibfield  {author} {\bibinfo {author} {\bibfnamefont {S.}~\bibnamefont {Mignuzzi}}, \bibinfo {author} {\bibfnamefont {S.}~\bibnamefont {Vezzoli}}, \bibinfo {author} {\bibfnamefont {S.~A.~R.}\ \bibnamefont {Horsley}}, \bibinfo {author} {\bibfnamefont {W.~L.}\ \bibnamefont {Barnes}}, \bibinfo {author} {\bibfnamefont {S.~A.}\ \bibnamefont {Maier}},\ and\ \bibinfo {author} {\bibfnamefont {R.}~\bibnamefont {Sapienza}},\ }\bibfield  {title} {\bibinfo {title} {Nanoscale design of the local density of optical states},\ }\href {https://doi.org/10.1021/acs.nanolett.8b04515} {\bibfield  {journal} {\bibinfo  {journal} {Nano Letters}\ }\textbf {\bibinfo {volume} {19}},\ \bibinfo {pages} {1613} (\bibinfo {year} {2019})}\BibitemShut {NoStop}%
\bibitem [{\citenamefont {Bennett}\ and\ \citenamefont {Buhmann}(2020)}]{Bennett2020}%
  \BibitemOpen
  \bibfield  {author} {\bibinfo {author} {\bibfnamefont {R.}~\bibnamefont {Bennett}}\ and\ \bibinfo {author} {\bibfnamefont {S.~Y.}\ \bibnamefont {Buhmann}},\ }\bibfield  {title} {\bibinfo {title} {Inverse design of light–matter interactions in macroscopic qed},\ }\href {https://doi.org/10.1088/1367-2630/abac3a} {\bibfield  {journal} {\bibinfo  {journal} {New Journal of Physics}\ }\textbf {\bibinfo {volume} {22}},\ \bibinfo {pages} {093014} (\bibinfo {year} {2020})}\BibitemShut {NoStop}%
\bibitem [{\citenamefont {Bennett}(2021)}]{Bennett2021}%
  \BibitemOpen
  \bibfield  {author} {\bibinfo {author} {\bibfnamefont {R.}~\bibnamefont {Bennett}},\ }\bibfield  {title} {\bibinfo {title} {Inverse design of environment-induced coherence},\ }\href {https://doi.org/10.1103/PhysRevA.103.013706} {\bibfield  {journal} {\bibinfo  {journal} {Phys. Rev. A}\ }\textbf {\bibinfo {volume} {103}},\ \bibinfo {pages} {013706} (\bibinfo {year} {2021})}\BibitemShut {NoStop}%
\bibitem [{\citenamefont {Jensen}\ and\ \citenamefont {Sigmund}(2011)}]{Jensen2011}%
  \BibitemOpen
  \bibfield  {author} {\bibinfo {author} {\bibfnamefont {J.}~\bibnamefont {Jensen}}\ and\ \bibinfo {author} {\bibfnamefont {O.}~\bibnamefont {Sigmund}},\ }\bibfield  {title} {\bibinfo {title} {Topology optimization for nano-photonics},\ }\href {https://doi.org/https://doi.org/10.1002/lpor.201000014} {\bibfield  {journal} {\bibinfo  {journal} {Laser \& Photonics Reviews}\ }\textbf {\bibinfo {volume} {5}},\ \bibinfo {pages} {308} (\bibinfo {year} {2011})}\BibitemShut {NoStop}%
\bibitem [{\citenamefont {Miguel-Torcal}\ \emph {et~al.}(2022)\citenamefont {Miguel-Torcal}, \citenamefont {Abad-Arredondo}, \citenamefont {García-Vidal},\ and\ \citenamefont {Fernández-Domínguez}}]{AMT2022}%
  \BibitemOpen
  \bibfield  {author} {\bibinfo {author} {\bibfnamefont {A.}~\bibnamefont {Miguel-Torcal}}, \bibinfo {author} {\bibfnamefont {J.}~\bibnamefont {Abad-Arredondo}}, \bibinfo {author} {\bibfnamefont {F.~J.}\ \bibnamefont {García-Vidal}},\ and\ \bibinfo {author} {\bibfnamefont {A.~I.}\ \bibnamefont {Fernández-Domínguez}},\ }\bibfield  {title} {\bibinfo {title} {Inverse-designed dielectric cloaks for entanglement generation},\ }\href {https://doi.org/doi:10.1515/nanoph-2022-0231} {\bibfield  {journal} {\bibinfo  {journal} {Nanophotonics}\ }\textbf {\bibinfo {volume} {11}},\ \bibinfo {pages} {4387} (\bibinfo {year} {2022})}\BibitemShut {NoStop}%
\bibitem [{\citenamefont {Miguel-Torcal}\ \emph {et~al.}(2024)\citenamefont {Miguel-Torcal}, \citenamefont {Gonz\'{a}lez-Tudela}, \citenamefont {Garc\'{i}a-Vidal},\ and\ \citenamefont {Fern\'{a}ndez-Dom\'{i}nguez}}]{AMT2024}%
  \BibitemOpen
  \bibfield  {author} {\bibinfo {author} {\bibfnamefont {A.}~\bibnamefont {Miguel-Torcal}}, \bibinfo {author} {\bibfnamefont {A.}~\bibnamefont {Gonz\'{a}lez-Tudela}}, \bibinfo {author} {\bibfnamefont {F.~J.}\ \bibnamefont {Garc\'{i}a-Vidal}},\ and\ \bibinfo {author} {\bibfnamefont {A.~I.}\ \bibnamefont {Fern\'{a}ndez-Dom\'{i}nguez}},\ }\bibfield  {title} {\bibinfo {title} {Multiqubit quantum state preparation enabled by topology optimization},\ }\href {https://doi.org/10.1364/OPTICAQ.530865} {\bibfield  {journal} {\bibinfo  {journal} {Optica Quantum}\ }\textbf {\bibinfo {volume} {2}},\ \bibinfo {pages} {371} (\bibinfo {year} {2024})}\BibitemShut {NoStop}%
\bibitem [{\citenamefont {Chakravarthi}\ \emph {et~al.}(2020)\citenamefont {Chakravarthi}, \citenamefont {Chao}, \citenamefont {Pederson}, \citenamefont {Molesky}, \citenamefont {Ivanov}, \citenamefont {Hestroffer}, \citenamefont {Hatami}, \citenamefont {Rodriguez},\ and\ \citenamefont {Fu}}]{Chakravarthi2020}%
  \BibitemOpen
  \bibfield  {author} {\bibinfo {author} {\bibfnamefont {S.}~\bibnamefont {Chakravarthi}}, \bibinfo {author} {\bibfnamefont {P.}~\bibnamefont {Chao}}, \bibinfo {author} {\bibfnamefont {C.}~\bibnamefont {Pederson}}, \bibinfo {author} {\bibfnamefont {S.}~\bibnamefont {Molesky}}, \bibinfo {author} {\bibfnamefont {A.}~\bibnamefont {Ivanov}}, \bibinfo {author} {\bibfnamefont {K.}~\bibnamefont {Hestroffer}}, \bibinfo {author} {\bibfnamefont {F.}~\bibnamefont {Hatami}}, \bibinfo {author} {\bibfnamefont {A.~W.}\ \bibnamefont {Rodriguez}},\ and\ \bibinfo {author} {\bibfnamefont {K.-M.~C.}\ \bibnamefont {Fu}},\ }\bibfield  {title} {\bibinfo {title} {Inverse-designed photon extractors for optically addressable defect qubits},\ }\href {https://doi.org/10.1364/OPTICA.408611} {\bibfield  {journal} {\bibinfo  {journal} {Optica}\ }\textbf {\bibinfo {volume} {7}},\ \bibinfo {pages} {1805} (\bibinfo {year} {2020})}\BibitemShut {NoStop}%
\bibitem [{\citenamefont {Melo}\ \emph {et~al.}(2023)\citenamefont {Melo}, \citenamefont {Eshbaugh}, \citenamefont {Flagg},\ and\ \citenamefont {Davanco}}]{Melo2023}%
  \BibitemOpen
  \bibfield  {author} {\bibinfo {author} {\bibfnamefont {E.~G.}\ \bibnamefont {Melo}}, \bibinfo {author} {\bibfnamefont {W.}~\bibnamefont {Eshbaugh}}, \bibinfo {author} {\bibfnamefont {E.~B.}\ \bibnamefont {Flagg}},\ and\ \bibinfo {author} {\bibfnamefont {M.}~\bibnamefont {Davanco}},\ }\bibfield  {title} {\bibinfo {title} {Multiobjective inverse design of solid-state quantum emitter single-photon sources},\ }\href {https://doi.org/10.1021/acsphotonics.2c00929} {\bibfield  {journal} {\bibinfo  {journal} {ACS Photonics}\ }\textbf {\bibinfo {volume} {10}},\ \bibinfo {pages} {959} (\bibinfo {year} {2023})}\BibitemShut {NoStop}%
\bibitem [{\citenamefont {Christiansen}\ \emph {et~al.}(2019)\citenamefont {Christiansen}, \citenamefont {Wang},\ and\ \citenamefont {Sigmund}}]{Sigmund2019}%
  \BibitemOpen
  \bibfield  {author} {\bibinfo {author} {\bibfnamefont {R.~E.}\ \bibnamefont {Christiansen}}, \bibinfo {author} {\bibfnamefont {F.}~\bibnamefont {Wang}},\ and\ \bibinfo {author} {\bibfnamefont {O.}~\bibnamefont {Sigmund}},\ }\bibfield  {title} {\bibinfo {title} {Topological insulators by topology optimization},\ }\href {https://doi.org/10.1103/PhysRevLett.122.234502} {\bibfield  {journal} {\bibinfo  {journal} {Phys. Rev. Lett.}\ }\textbf {\bibinfo {volume} {122}},\ \bibinfo {pages} {234502} (\bibinfo {year} {2019})}\BibitemShut {NoStop}%
\bibitem [{\citenamefont {Chen}\ \emph {et~al.}(2020)\citenamefont {Chen}, \citenamefont {Meng}, \citenamefont {Kivshar}, \citenamefont {Jia},\ and\ \citenamefont {Huang}}]{Chen2020}%
  \BibitemOpen
  \bibfield  {author} {\bibinfo {author} {\bibfnamefont {Y.}~\bibnamefont {Chen}}, \bibinfo {author} {\bibfnamefont {F.}~\bibnamefont {Meng}}, \bibinfo {author} {\bibfnamefont {Y.}~\bibnamefont {Kivshar}}, \bibinfo {author} {\bibfnamefont {B.}~\bibnamefont {Jia}},\ and\ \bibinfo {author} {\bibfnamefont {X.}~\bibnamefont {Huang}},\ }\bibfield  {title} {\bibinfo {title} {Inverse design of higher-order photonic topological insulators},\ }\href {https://doi.org/10.1103/PhysRevResearch.2.023115} {\bibfield  {journal} {\bibinfo  {journal} {Phys. Rev. Res.}\ }\textbf {\bibinfo {volume} {2}},\ \bibinfo {pages} {023115} (\bibinfo {year} {2020})}\BibitemShut {NoStop}%
\bibitem [{\citenamefont {Su}\ \emph {et~al.}(1979)\citenamefont {Su}, \citenamefont {Schrieffer},\ and\ \citenamefont {Heeger}}]{SSH1979}%
  \BibitemOpen
  \bibfield  {author} {\bibinfo {author} {\bibfnamefont {W.~P.}\ \bibnamefont {Su}}, \bibinfo {author} {\bibfnamefont {J.~R.}\ \bibnamefont {Schrieffer}},\ and\ \bibinfo {author} {\bibfnamefont {A.~J.}\ \bibnamefont {Heeger}},\ }\bibfield  {title} {\bibinfo {title} {Solitons in polyacetylene},\ }\href {https://doi.org/10.1103/PhysRevLett.42.1698} {\bibfield  {journal} {\bibinfo  {journal} {Phys. Rev. Lett.}\ }\textbf {\bibinfo {volume} {42}},\ \bibinfo {pages} {1698} (\bibinfo {year} {1979})}\BibitemShut {NoStop}%
\bibitem [{\citenamefont {Pocock}\ \emph {et~al.}(2018)\citenamefont {Pocock}, \citenamefont {Xiao}, \citenamefont {Huidobro},\ and\ \citenamefont {Giannini}}]{Giannini2018}%
  \BibitemOpen
  \bibfield  {author} {\bibinfo {author} {\bibfnamefont {S.~R.}\ \bibnamefont {Pocock}}, \bibinfo {author} {\bibfnamefont {X.}~\bibnamefont {Xiao}}, \bibinfo {author} {\bibfnamefont {P.~A.}\ \bibnamefont {Huidobro}},\ and\ \bibinfo {author} {\bibfnamefont {V.}~\bibnamefont {Giannini}},\ }\bibfield  {title} {\bibinfo {title} {Topological plasmonic chain with retardation and radiative effects},\ }\href {https://doi.org/10.1021/acsphotonics.8b00117} {\bibfield  {journal} {\bibinfo  {journal} {ACS Photonics}\ }\textbf {\bibinfo {volume} {5}},\ \bibinfo {pages} {2271} (\bibinfo {year} {2018})}\BibitemShut {NoStop}%
\bibitem [{\citenamefont {Buendía}\ \emph {et~al.}(2023)\citenamefont {Buendía}, \citenamefont {Sánchez-Gil},\ and\ \citenamefont {Giannini}}]{Buendia2023}%
  \BibitemOpen
  \bibfield  {author} {\bibinfo {author} {\bibfnamefont {A.}~\bibnamefont {Buendía}}, \bibinfo {author} {\bibfnamefont {J.~A.}\ \bibnamefont {Sánchez-Gil}},\ and\ \bibinfo {author} {\bibfnamefont {V.}~\bibnamefont {Giannini}},\ }\bibfield  {title} {\bibinfo {title} {Exploiting oriented field projectors to open topological gaps in plasmonic nanoparticle arrays},\ }\href {https://doi.org/10.1021/acsphotonics.2c01526} {\bibfield  {journal} {\bibinfo  {journal} {ACS Photonics}\ }\textbf {\bibinfo {volume} {10}},\ \bibinfo {pages} {464} (\bibinfo {year} {2023})}\BibitemShut {NoStop}%
\bibitem [{\citenamefont {Dung}\ \emph {et~al.}(2002)\citenamefont {Dung}, \citenamefont {Kn\"oll},\ and\ \citenamefont {Welsch}}]{Dung2002}%
  \BibitemOpen
  \bibfield  {author} {\bibinfo {author} {\bibfnamefont {H.~T.}\ \bibnamefont {Dung}}, \bibinfo {author} {\bibfnamefont {L.}~\bibnamefont {Kn\"oll}},\ and\ \bibinfo {author} {\bibfnamefont {D.-G.}\ \bibnamefont {Welsch}},\ }\bibfield  {title} {\bibinfo {title} {Resonant dipole-dipole interaction in the presence of dispersing and absorbing surroundings},\ }\href {https://doi.org/10.1103/PhysRevA.66.063810} {\bibfield  {journal} {\bibinfo  {journal} {Phys. Rev. A}\ }\textbf {\bibinfo {volume} {66}},\ \bibinfo {pages} {063810} (\bibinfo {year} {2002})}\BibitemShut {NoStop}%
\bibitem [{\citenamefont {Breuer}\ and\ \citenamefont {Petruccione}(2007)}]{Petruccione2007}%
  \BibitemOpen
  \bibfield  {author} {\bibinfo {author} {\bibfnamefont {H.-P.}\ \bibnamefont {Breuer}}\ and\ \bibinfo {author} {\bibfnamefont {F.}~\bibnamefont {Petruccione}},\ }\href {https://doi.org/10.1093/acprof:oso/9780199213900.001.0001} {\emph {\bibinfo {title} {{The Theory of Open Quantum Systems}}}}\ (\bibinfo  {publisher} {Oxford University Press},\ \bibinfo {year} {2007})\BibitemShut {NoStop}%
\bibitem [{\citenamefont {Novotny}\ and\ \citenamefont {Hecht}(2012)}]{Novotny2012}%
  \BibitemOpen
  \bibfield  {author} {\bibinfo {author} {\bibfnamefont {L.}~\bibnamefont {Novotny}}\ and\ \bibinfo {author} {\bibfnamefont {B.}~\bibnamefont {Hecht}},\ }\href {https://doi.org/10.1017/CBO9780511794193} {\emph {\bibinfo {title} {Principles of Nano-Optics}}},\ \bibinfo {edition} {2nd}\ ed.\ (\bibinfo  {publisher} {Cambridge University Press},\ \bibinfo {year} {2012})\BibitemShut {NoStop}%
\bibitem [{\citenamefont {Gonzalez-Tudela}\ \emph {et~al.}(2011)\citenamefont {Gonzalez-Tudela}, \citenamefont {Martin-Cano}, \citenamefont {Moreno}, \citenamefont {Martin-Moreno}, \citenamefont {Tejedor},\ and\ \citenamefont {Garcia-Vidal}}]{GonzalezTudela2011}%
  \BibitemOpen
  \bibfield  {author} {\bibinfo {author} {\bibfnamefont {A.}~\bibnamefont {Gonzalez-Tudela}}, \bibinfo {author} {\bibfnamefont {D.}~\bibnamefont {Martin-Cano}}, \bibinfo {author} {\bibfnamefont {E.}~\bibnamefont {Moreno}}, \bibinfo {author} {\bibfnamefont {L.}~\bibnamefont {Martin-Moreno}}, \bibinfo {author} {\bibfnamefont {C.}~\bibnamefont {Tejedor}},\ and\ \bibinfo {author} {\bibfnamefont {F.~J.}\ \bibnamefont {Garcia-Vidal}},\ }\bibfield  {title} {\bibinfo {title} {Entanglement of two qubits mediated by one-dimensional plasmonic waveguides},\ }\href {https://doi.org/10.1103/PhysRevLett.106.020501} {\bibfield  {journal} {\bibinfo  {journal} {Phys. Rev. Lett.}\ }\textbf {\bibinfo {volume} {106}},\ \bibinfo {pages} {020501} (\bibinfo {year} {2011})}\BibitemShut {NoStop}%
\bibitem [{\citenamefont {Eliseev}\ \emph {et~al.}(2000)\citenamefont {Eliseev}, \citenamefont {Li}, \citenamefont {Stintz}, \citenamefont {Liu}, \citenamefont {Newell}, \citenamefont {Malloy},\ and\ \citenamefont {Lester}}]{Eliseev2000}%
  \BibitemOpen
  \bibfield  {author} {\bibinfo {author} {\bibfnamefont {P.~G.}\ \bibnamefont {Eliseev}}, \bibinfo {author} {\bibfnamefont {H.}~\bibnamefont {Li}}, \bibinfo {author} {\bibfnamefont {A.}~\bibnamefont {Stintz}}, \bibinfo {author} {\bibfnamefont {G.~T.}\ \bibnamefont {Liu}}, \bibinfo {author} {\bibfnamefont {T.~C.}\ \bibnamefont {Newell}}, \bibinfo {author} {\bibfnamefont {K.~J.}\ \bibnamefont {Malloy}},\ and\ \bibinfo {author} {\bibfnamefont {L.~F.}\ \bibnamefont {Lester}},\ }\bibfield  {title} {\bibinfo {title} {Transition dipole moment of inas/ingaas quantum dots from experiments on ultralow-threshold laser diodes},\ }\href {https://doi.org/10.1063/1.126944} {\bibfield  {journal} {\bibinfo  {journal} {Applied Physics Letters}\ }\textbf {\bibinfo {volume} {77}},\ \bibinfo {pages} {262} (\bibinfo {year} {2000})}\BibitemShut {NoStop}%
\bibitem [{\citenamefont {Vivas-Via\~na}\ and\ \citenamefont {S\'anchez Mu\~noz}(2021)}]{Vivas2021}%
  \BibitemOpen
  \bibfield  {author} {\bibinfo {author} {\bibfnamefont {A.}~\bibnamefont {Vivas-Via\~na}}\ and\ \bibinfo {author} {\bibfnamefont {C.}~\bibnamefont {S\'anchez Mu\~noz}},\ }\bibfield  {title} {\bibinfo {title} {Two-photon resonance fluorescence of two interacting nonidentical quantum emitters},\ }\href {https://doi.org/10.1103/PhysRevResearch.3.033136} {\bibfield  {journal} {\bibinfo  {journal} {Phys. Rev. Res.}\ }\textbf {\bibinfo {volume} {3}},\ \bibinfo {pages} {033136} (\bibinfo {year} {2021})}\BibitemShut {NoStop}%
\bibitem [{\citenamefont {Ficek}\ and\ \citenamefont {Tanaś}(2002)}]{Ficek2002}%
  \BibitemOpen
  \bibfield  {author} {\bibinfo {author} {\bibfnamefont {Z.}~\bibnamefont {Ficek}}\ and\ \bibinfo {author} {\bibfnamefont {R.}~\bibnamefont {Tanaś}},\ }\bibfield  {title} {\bibinfo {title} {Entangled states and collective nonclassical effects in two-atom systems},\ }\href {https://doi.org/https://doi.org/10.1016/S0370-1573(02)00368-X} {\bibfield  {journal} {\bibinfo  {journal} {Physics Reports}\ }\textbf {\bibinfo {volume} {372}},\ \bibinfo {pages} {369} (\bibinfo {year} {2002})}\BibitemShut {NoStop}%
\bibitem [{\citenamefont {Asbóth}\ \emph {et~al.}(2016)\citenamefont {Asbóth}, \citenamefont {Oroszlány},\ and\ \citenamefont {Pályi}}]{Asboth2016}%
  \BibitemOpen
  \bibfield  {author} {\bibinfo {author} {\bibfnamefont {J.}~\bibnamefont {Asbóth}}, \bibinfo {author} {\bibfnamefont {L.}~\bibnamefont {Oroszlány}},\ and\ \bibinfo {author} {\bibfnamefont {A.}~\bibnamefont {Pályi}},\ }\href {https://doi.org/https://doi.org/10.1007/978-3-319-25607-8} {\emph {\bibinfo {title} {A short course on topological insulators}}},\ \bibinfo {edition} {1st}\ ed.\ (\bibinfo  {publisher} {Springer Cham},\ \bibinfo {year} {2016})\BibitemShut {NoStop}%
\bibitem [{\citenamefont {Ando}(2013)}]{Ando2013}%
  \BibitemOpen
  \bibfield  {author} {\bibinfo {author} {\bibfnamefont {Y.}~\bibnamefont {Ando}},\ }\bibfield  {title} {\bibinfo {title} {Topological insulator materials},\ }\href {https://doi.org/10.7566/JPSJ.82.102001} {\bibfield  {journal} {\bibinfo  {journal} {Journal of the Physical Society of Japan}\ }\textbf {\bibinfo {volume} {82}},\ \bibinfo {pages} {102001} (\bibinfo {year} {2013})}\BibitemShut {NoStop}%
\bibitem [{\citenamefont {Deng}\ \emph {et~al.}(2021)\citenamefont {Deng}, \citenamefont {Pan}, \citenamefont {Chen},\ and\ \citenamefont {Zhai}}]{Deng2021}%
  \BibitemOpen
  \bibfield  {author} {\bibinfo {author} {\bibfnamefont {T.-S.}\ \bibnamefont {Deng}}, \bibinfo {author} {\bibfnamefont {L.}~\bibnamefont {Pan}}, \bibinfo {author} {\bibfnamefont {Y.}~\bibnamefont {Chen}},\ and\ \bibinfo {author} {\bibfnamefont {H.}~\bibnamefont {Zhai}},\ }\bibfield  {title} {\bibinfo {title} {Stability of time-reversal symmetry protected topological phases},\ }\href {https://doi.org/10.1103/PhysRevLett.127.086801} {\bibfield  {journal} {\bibinfo  {journal} {Phys. Rev. Lett.}\ }\textbf {\bibinfo {volume} {127}},\ \bibinfo {pages} {086801} (\bibinfo {year} {2021})}\BibitemShut {NoStop}%
\bibitem [{\citenamefont {Nie}\ \emph {et~al.}(2021{\natexlab{b}})\citenamefont {Nie}, \citenamefont {Shi}, \citenamefont {Nori},\ and\ \citenamefont {Liu}}]{Nie2021}%
  \BibitemOpen
  \bibfield  {author} {\bibinfo {author} {\bibfnamefont {W.}~\bibnamefont {Nie}}, \bibinfo {author} {\bibfnamefont {T.}~\bibnamefont {Shi}}, \bibinfo {author} {\bibfnamefont {F.}~\bibnamefont {Nori}},\ and\ \bibinfo {author} {\bibfnamefont {Y.-x.}\ \bibnamefont {Liu}},\ }\bibfield  {title} {\bibinfo {title} {Topology-enhanced nonreciprocal scattering and photon absorption in a waveguide},\ }\href {https://doi.org/10.1103/PhysRevApplied.15.044041} {\bibfield  {journal} {\bibinfo  {journal} {Phys. Rev. Appl.}\ }\textbf {\bibinfo {volume} {15}},\ \bibinfo {pages} {044041} (\bibinfo {year} {2021}{\natexlab{b}})}\BibitemShut {NoStop}%
\bibitem [{\citenamefont {Zak}(1989)}]{Zak1989}%
  \BibitemOpen
  \bibfield  {author} {\bibinfo {author} {\bibfnamefont {J.}~\bibnamefont {Zak}},\ }\bibfield  {title} {\bibinfo {title} {Berry's phase for energy bands in solids},\ }\href {https://doi.org/10.1103/PhysRevLett.62.2747} {\bibfield  {journal} {\bibinfo  {journal} {Phys. Rev. Lett.}\ }\textbf {\bibinfo {volume} {62}},\ \bibinfo {pages} {2747} (\bibinfo {year} {1989})}\BibitemShut {NoStop}%
\bibitem [{\citenamefont {Chen}\ and\ \citenamefont {Chiou}(2020)}]{Chiou2020}%
  \BibitemOpen
  \bibfield  {author} {\bibinfo {author} {\bibfnamefont {B.-H.}\ \bibnamefont {Chen}}\ and\ \bibinfo {author} {\bibfnamefont {D.-W.}\ \bibnamefont {Chiou}},\ }\bibfield  {title} {\bibinfo {title} {An elementary rigorous proof of bulk-boundary correspondence in the generalized su-schrieffer-heeger model},\ }\href {https://doi.org/https://doi.org/10.1016/j.physleta.2019.126168} {\bibfield  {journal} {\bibinfo  {journal} {Physics Letters A}\ }\textbf {\bibinfo {volume} {384}},\ \bibinfo {pages} {126168} (\bibinfo {year} {2020})}\BibitemShut {NoStop}%
\bibitem [{\citenamefont {McDonnell}\ and\ \citenamefont {Olmos}(2022)}]{McDonnell2022}%
  \BibitemOpen
  \bibfield  {author} {\bibinfo {author} {\bibfnamefont {C.}~\bibnamefont {McDonnell}}\ and\ \bibinfo {author} {\bibfnamefont {B.}~\bibnamefont {Olmos}},\ }\bibfield  {title} {\bibinfo {title} {Subradiant edge states in an atom chain with waveguide-mediated hopping},\ }\href {https://doi.org/10.22331/q-2022-09-15-805} {\bibfield  {journal} {\bibinfo  {journal} {{Quantum}}\ }\textbf {\bibinfo {volume} {6}},\ \bibinfo {pages} {805} (\bibinfo {year} {2022})}\BibitemShut {NoStop}%
\bibitem [{\citenamefont {Jozsa}(1994)}]{Jozsa1994}%
  \BibitemOpen
  \bibfield  {author} {\bibinfo {author} {\bibfnamefont {R.}~\bibnamefont {Jozsa}},\ }\bibfield  {title} {\bibinfo {title} {Fidelity for mixed quantum states},\ }\href {https://doi.org/10.1080/09500349414552171} {\bibfield  {journal} {\bibinfo  {journal} {Journal of Modern Optics}\ }\textbf {\bibinfo {volume} {41}},\ \bibinfo {pages} {2315} (\bibinfo {year} {1994})}\BibitemShut {NoStop}%
\end{thebibliography}%

\end{document}